\documentclass[aps,prb,twocolumn,showpacs,tightenlines,amsmath]{revtex4}
 \usepackage{graphics}
 \usepackage{epsfig}
   \begin{document} 
   \title{\Large \bf{
           Floquet scattering theory for
           current and heat noise 
           in large amplitude
           adiabatic pumps
                    }
         }
\author{
M. Moskalets$^{1}$
and
M. B\"uttiker$^2$
}
\affiliation{
     $^1$Department of Metal and Semiconductor Physics,\\
     National Technical University "Kharkiv Polytechnic Institute",
     61002 Kharkiv, Ukraine\\
     $^2$D\'epartement de Physique Th\'eorique, Universit\'e de Gen\`eve,
     CH-1211 Gen\`eve 4, Switzerland\\}
\date\today
   \begin{abstract}
We discuss the statistical correlation properties of currents and
energy flows generated by an adiabatic quantum pump. 
Our approach emphasizes the important role of quantized energy exchange 
between the sea of electrons and the oscillating scatterer. 
The frequency $\omega$ of oscillations introduces a natural energy
scale $\hbar\omega$. 
In the low temperature limit $k_BT\ll\hbar\omega$ 
the pump generates a shot-like noise which 
manifests itself in photon-assisted quantum mechanical exchange amplitudes. 
In the high temperature limit $k_BT\gg\hbar\omega$ the pump produces 
a thermal-like noise due to ac-currents generated by the pump. 
We predict that with increasing temperature the frequency dependence
of the noise changes.
The current noise is linear in $\omega$ at low temperatures, is 
quadratic at intermediate temperatures, and is linear again at high temperatures.
Similarly, in the same temperature regions, the heat flow noise is proportional to $\omega^3$,  
$\omega^2$, and $\omega$.
   \end{abstract}
\maketitle
\small

\section{Introduction}
\label{intro}

Quantum pumping, a phenomenon in which a
periodic local perturbation gives rise to a directed current in a phase
coherent mesoscopic system, attracts great attention of both 
experimentalists 
%\cite{SMCG99,HLWB01,WPMU03,DMH03} 
\cite{SMCG99}$^{-}$\cite{DMH03} 
and theorists
%\cite{Thouless83,BTP94,Brouwer98,AA98,ZSA99,
%SAA00,AEGS00,SW00,VAA01,AEGS01,TC01,MB01,BH01,
%WWG02,CB02,ZW02,EWAL02,MBstrong02,Kim02,ZCMcK03,
%GTM03,PB02,AEGSS02,AEGS03,ZLCMcK03,CTCC03,RS03,FHP03,TB04}.
\cite{Thouless83}$^{-}$\cite{TB04}.
The physical mechanism leading to adiabatic quantum pumping
involves quantum-mechanical interference and
dynamical breaking of time-reversal invariance. 
This mechanism is relevant not only for open (i.e., connected to
external particle reservoirs) systems but also for closed (ring-like)
mesoscopic systems. \cite{Cohen02,MB03,CKS04}

The possibility 
%\cite{Thouless83,AEGS01,WWG00,LEWW01,WWGR01,
%WWq02,EWA02,EWAK02,BDR03,KAEW03} 
\cite{Thouless83,AEGS01,WWG00}$^{-}$\cite{KAEW03}
to achieve quantized transport,
not only of charge but in addition heat
\cite{AEGS01,MB02,WW02,MBstrong02} and spin
%\cite{WPMU03,SC01,MCM01,WWW02,TBB02,GTF02,
%Aono03,ZWWWSG03,BB03,WWG03,SB03,CAN03,BB04,BT04,SO04} 
\cite{WPMU03,SC01}$^{-}$\cite{SO04}
currents, makes pumping interesting also in view of 
possible applications. Quantum pumping has been investigated 
in systems of strongly correlated electrons
\cite{AA98,SC01,WWKondo02,CAN03}, for systems 
in the quantum Hall regime\cite{Simon00,Blaauboer03} and 
in hybrid superconducting-normal structures.
%\cite{Zhou99,WWWG01,JWBW02,Blaauboer02,WeiW02,BWJW02,TGF04}
\cite{Zhou99}$^{-}$\cite{TGF04}
Clearly any mesoscopic system with periodically evolving
properties is able to exhibit a quantum pump effect.

\begin{figure}[b]
  \vspace{3mm}
  \centerline{
   \epsfxsize8cm
 \epsffile{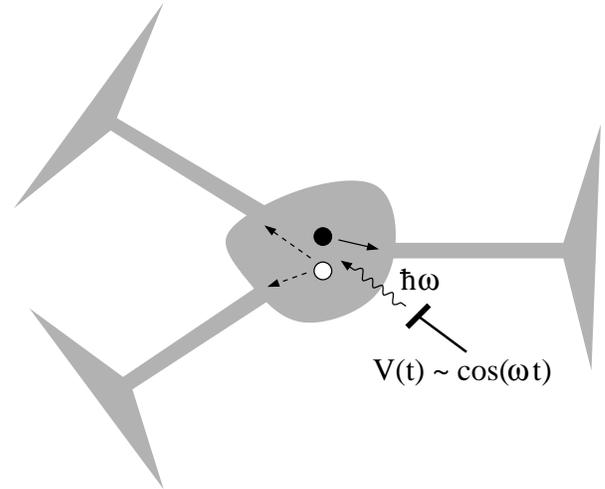}
             }
  \vspace{3mm}
  \nopagebreak
  \caption{ 
Noise of a quantum pump:
Two particles, an electron (open circle) and a hole (black circle),
are involved in the scattering process relevant for the  
quantum noise.
Different final states of a particle are possible: a particle 
can be transmitted into either of the leads (shown by dashed 
lines). These processes are described by photon-assisted quantum-mechanical exchange amplitudes.
   }
\label{fig0}
\end{figure}

Since the pump works under the influence of a time-dependent
(periodic) perturbation it generates time-dependent (ac) currents.
\cite{BTP94,MBac03} The dc current, which is mainly of
interest, is just the time averaged fully time-dependent current
generated by the pump.
The ac currents manifest themselves, for
instance,  in an interference effect \cite{MBac03}
with ac currents driven 
through the pump if an external ac bias
%\cite{DMH03,Brouwer01,PB01,MMLM03} 
\cite{DMH03,Brouwer01}$^{-}$\cite{MMLM03}
is applied. In addition, as we
will show in this work, the ac currents are visible in the noise of an
unbiased pump, Fig.\ref{fig0}.

The noise of a pump is important
because it is closely related to whether quantized pumping is possible.
\cite{AEGS01,AK00,MM01}
In addition the noise contains information
about the physical processes taking place in quantum pumps which 
can not be obtained by considering only the time-averaged pump current.
%\cite{PB02,Levitov01,MB02,PVB02,MA03,CLKH03,CKH04}
\cite{PB02,Levitov01,MB02}$^{-}$\cite{CKH04}

In the present paper we use the Floquet scattering matrix approach
%\cite{Shirley65,mbrl,Wagner94,PB98,MBstrong02,GH98,GA03}
\cite{Shirley65}$^{-}$\cite{GA03}
to investigate the quantum statistical correlation properties (noise) 
of multi-terminal adiabatic quantum pumps, Fig.\ref{fig0}. 
Our approach is based on the scattering matrix approach to ac transport in  
phase coherent mesoscopic systems \cite{BTP94}.  
According to this approach the currents flowing in the system  
are determined by the scattering  of electrons coming from the  
reservoirs by the mesoscopic sample \cite{Buttiker90,Buttiker92}. 
The basics of the approach used here is presented in 
Refs.~\onlinecite{MBstrong02,MBac03} and 
the results obtained here generalize
Ref.~\onlinecite{MB02} to the case of a large amplitude pump.

Starting from a general formalism we mainly deal with the low frequency limit 
that differentiates our work from other works employing the Floquet approach to 
a current noise problem and concentrating on a limit of high driving frequencies
(see, e.g., Refs.~\onlinecite{CLKH03,CKH04}).
The low frequency limit corresponds to {\it adiabatic} quantum pumping
which was investigated experimentally in Ref.~\onlinecite{SMCG99}.

The Floquet scattering matrix approach,
on the one hand, emphasizes the existence of side bands of 
electrons exiting the pump: these side bands are directly connected to the ac currents
generated by the pump. 
On the other hand, this approach allows to reformulate an intrinsically  
non-stationary problem in terms of a stationary one. 
The latter is more convenient for the investigation reported here, for instance, it permits a transparent discussion of 
heat flow and its fluctuations. 
Therefore within the framework of the Floquet scattering matrix approach 
we can consider two aspects of particle flow fluctuations on the same 
footing: 
(i) fluctuations of the charge (current) flow, and 
(ii) fluctuations of the energy (heat) flow. 
We investigate these fluctuations and their dependence on the pump frequency 
at low and at relatively high temperatures.

The paper is organized as follows. 
In Sec.\ref{FS} we derive the general expressions for the current 
and heat flow noise in terms of the Floquet scattering matrix elements, 
and outline the adiabatic approximation used in the following sections. 
In Secs.\ref{ACNP} and \ref{AHFF} we calculate the current noise and
the heat flow noise, respectively, generated by an adiabatic 
quantum pump. 
In Sec.\ref{SE} we apply the general expressions of previous sections 
to a pump consisting of two oscillating delta-function barriers. 
We conclude in Sec.\ref{DC}.

\section{Floquet scattering matrix expressions for noise}
\label{FS}

We consider a  mesoscopic sample (scatterer) connected to $N_{r}$
reservoirs via single channel leads. The reservoirs are assumed to
to be at equal temperatures $T_{\alpha}$ and electrochemical
potentials $\mu_{\alpha}$
\begin{equation}
\label{Eq1}
\mu_{\alpha} = \mu; \quad T_{\alpha} = T, \quad \alpha = 1,\dots,N_r.
\end{equation}

Let $\hat a_{\alpha}(E)$ and $\hat b_{\alpha}(E)$ be the
annihilation operators for particles incident from and outgoing to
the reservoir $\alpha$, respectively. The operators $\hat
a_{\alpha}(E)$ for incoming particles obey the following anti
commutation relation
\begin{equation}
\label{Eq2}
 [\hat a^{\dagger}_{\alpha}(E), \hat a_{\beta}(E')] =
\delta_{\alpha\beta}\delta(E - E').
\end{equation}

\noindent We assume that all the reservoirs are unaffected by the
coupling to the scatterer and are in the equilibrium state.

Consequently, in all the leads, the quantum statistical
average denoted by $\langle\dots\rangle$ of the incoming particles
are those of an electron system at equilibrium. In particular,
$\langle\hat a^{\dagger}_{\alpha}(E) \hat a_{\beta}(E')\rangle =
\delta_{\alpha\beta}\delta(E - E') f_{0}(E)$ is proportional to
the equilibrium Fermi distribution function,
\begin{equation}
\label{Eq3}
 f_{0}(E) = \frac{1}{\exp\left(\frac{E - \mu}{k_BT} \right) +1}.
\end{equation}

\noindent
$k_B$ is the Boltzmann constant.

The operators $\hat b_{\alpha}(E)$ for outgoing particles are
related to the operators $\hat a_{\alpha}(E)$ through the
scattering matrix of the sample under consideration.
\cite{Buttiker92}

We consider a sample that is subject to external forces that are
periodic in time. In particular its scattering properties
oscillate in time with period ${\cal T} = 2\pi/\omega$. As a
consequence an electron interacting with a scatterer can gain or
loss one or several energy quanta $n\hbar\omega,~n = 0,\pm 1,\pm
2,\dots$. Scattering on such an oscillatory scatterer can be
described via the Floquet scattering matrix. We are interested in
the sub-matrix of the Floquet matrix which describes the
transitions between the propagating states. \cite{MBstrong02} We
denote this sub-matrix by $\hat S_{F}$. The elements
$S_{F,\alpha\beta}(E_n,E)$ of this matrix are the quantum
mechanical amplitudes for an electron with energy $E$ entering the
scatterer through lead $\beta$ and to leave the scatterer with
energy $E_n = E + n\hbar\omega$ through lead $\alpha$.  The
particle has thus absorbed ($n > 0$) or emitted ($n < 0$) energy
quanta $|n|\hbar\omega$. The matrix $\hat S_{F}$ is unitary:
\begin{subequations}
\label{Eq4}
\begin{equation}
\label{Eq4A}
\sum\limits_{\alpha}\sum\limits_{n=-\infty}^{\infty}
S^{*}_{F,\alpha\beta}(E_{n},E)S_{F,\alpha\gamma}(E_{n},E_{m})
= \delta_{m0}\delta_{\beta\gamma},
\end{equation}
\begin{equation}
\label{Eq4B}
\sum\limits_{\beta}\sum\limits_{n=-\infty}^{\infty}
S^{*}_{F,\alpha\beta}(E,E_{n})S_{F,\gamma\beta}(E_{m},E_{n})
= \delta_{m0}\delta_{\alpha\gamma}.
\end{equation}
\end{subequations}

Using the Floquet scattering matrix we can express the
annihilation operators for outgoing particles in terms of the
annihilation operators of the incoming particles,
\begin{equation}
\label{Eq5}
 \hat b_{\alpha}(E) = \sum_\beta \sum\limits_{n}
S_{F,\alpha\beta}(E,E_n) \hat a_{\beta}(E_n).
\end{equation}

The unitarity conditions Eqs. (\ref{Eq4}) guarantee that the
operators for outgoing particles obey the same anti-commutation
relations as operators for the incoming particles.

\subsection{Current noise}
\label{CN}

The correlation function ${\bf P}_{\alpha\beta}(t_{1}, t_{2})$
of currents is \cite{BB00} :
\begin{equation}
\label{Eq6}
{\bf P}_{\alpha\beta}(t_{1}, t_{2}) = \frac{1}{2}\langle
\Delta\hat I_{\alpha}(t_{1})\Delta\hat I_{\beta}(t_{2}) +
\Delta\hat I_{\beta}(t_{2})\Delta\hat I_{\alpha}(t_{1}) \rangle,
\end{equation}

\noindent Here $\hat I_{\alpha}(t)$ is the current operator in
lead $\alpha$ and $\Delta\hat I_{\alpha}(t) = \hat I_{\alpha}(t) -
\langle\hat I_{\alpha}(t)\rangle$. In the low temperature and low
frequency limit of interest here
\begin{equation}
\label{Eq7}
  k_BT, ~\hbar\omega \ll \mu,
\end{equation}

\noindent
the operator $\hat I_{\alpha}(t)$ is \cite{Buttiker92}
\begin{equation}
\label{Eq8}
\hat I_{\alpha}(t) = \frac{e}{h} \int dEdE^{\prime}
[\hat b^{\dagger}_{\alpha}(E)\hat b_{\alpha}(E^{\prime})
- \hat a^{\dagger}_{\alpha}(E)\hat a_{\alpha}(E^{\prime})]
e^{i\frac{E-E^{\prime}}{\hbar}t}.
\end{equation}

In the present work we investigate the correlation function for
currents generated by the oscillating scatterer (a quantum pump).
Since the pump is non-stationary the function ${\bf
P}_{\alpha\beta}(t_{1}, t_{2})$ depends on two times. However
the pump works slowly: the pump frequency  is small compared 
to the Fermi energy ($\omega\ll \mu/\hbar$) and it generates
only noise in states lying close to the zero frequency region. 
To make this noise visible we need to eliminate the contribution 
coming from high frequency quantum fluctuations. 
To this end we integrate the
correlation function ${\bf P}_{\alpha\beta}(t_{1}, t_{2})$ over
the time difference $\tau = t_{1} - t_{2}$
and get the function ${\bf P}_{\alpha\beta}(t)$ 
depending only on the middle time $t = (t_{1} + t_{2})/2$. 
This function is periodic in time with period ${\cal T}$ and it is
a correlation function of the currents generated by the pump.
\cite{MBac03}

We are interested in the zero-frequency Fourier coefficient
of this function.
Thus the quantity of interest is:
\begin{equation}
\label{Eq9}
{\bf P}_{\alpha\beta} = 2 \int\limits_{0}^{\cal T} \frac{dt}{{\cal T}}
\int\limits_{-\infty}^{\infty} d\tau
{\bf P}_{\alpha\beta}\Big(t + \frac{\tau}{2}, t - \frac{\tau}{2} \Big)
\end{equation}
The zero-frequency noise considered here corresponds to a noise
measurement which long compared to all intrinsic frequencies and
long compared to the pump frequency. We remark that for very slow
pumping cycles one could alternatively investigate the noise
spectrum for a particular value of  the pump parameters,
assuming that the noise measurement is carried out on a
times-scale short compared to the pump frequency.

Using Eqs. (\ref{Eq5}), (\ref{Eq8}) we can express ${\bf
P}_{\alpha\beta}$ in terms of the Floquet scattering matrix:
\begin{widetext}
\begin{subequations}
\label{Eq10}
\begin{equation}
\label{Eq10A}
{\bf P}_{\alpha\beta} = \frac{2e^2}{h}\int\limits_{0}^{\infty}dE
\Big( {\bf P}_{\alpha\beta}^{(th)}(E) + 
{\bf P}_{\alpha\beta}^{(sh)}(E) \Big) ,
\end{equation}
\begin{equation}
\label{Eq10B} {\bf P}_{\alpha\beta}^{(th)}(E) = f_0(E)[1 - f_0(E)]
\Big( \delta_{\alpha\beta} + \sum\limits_{n=-\infty}^{\infty}
\Big\{ \delta_{\alpha\beta} \sum\limits_{\gamma}
\big|S_{F,\alpha\gamma}(E_n,E)\big|^2  -
\big|S_{F,\alpha\beta}(E_n,E)\big|^2  -
\big|S_{F,\beta\alpha}(E_n,E)\big|^2 \Big\} \Big).
\end{equation}
\begin{equation}
\label{Eq10C} {\bf P}_{\alpha\beta}^{(sh)}(E) =
\sum\limits_{\gamma,\delta} \sum\limits_{n=-\infty}^{\infty}
\sum\limits_{m=-\infty}^{\infty} \sum\limits_{p=-\infty}^{\infty}
\frac{[f_0(E_n) - f_0(E_m) ]^2}{2} S_{F,\alpha\gamma}^{*}(E,E_n)
S_{F,\alpha\delta}(E,E_m) S_{F,\beta\delta}^{*}(E_p,E_m)
S_{F,\beta\gamma}(E_p,E_n).
\end{equation}
\end{subequations}
\end{widetext}

The first term ${\bf P}_{\alpha\beta}^{(th)}$ is thermal
(or Nyquist - Johnson) noise modified by the pump. It is due to 
thermal fluctuations of the incident states and 
disappears at zero temperature $T = 0$.

The second term ${\bf P}_{\alpha\beta}^{(sh)}$ is a shot noise
contribution. It is entirely due to a working pump. 
The terms in Eq. (\ref{Eq10C}) describe quantum mechanical 
exchange during
scattering of particles from the leads $\gamma$ and $\delta$ to
the leads $\alpha$ and $\beta$. The only difference from the
stationary case is that during a scattering process an electron can
change its energy absorbing or emitting energy quanta
$\hbar\omega$. The quanta emitted/absorbed are counted by 
$n$, $m$ and $p$ in Eqs. \ref{Eq10B} \ref{Eq10C}. In the scattering 
processes which contribute to the noise two
particles with energy $E_n$ and $E_m$ are incident through 
leads $\gamma$ and $\delta$, respectively. The particles leave the
scatterer through the lead $\alpha$ having energy $E$ and through
the lead $\beta$ having energy $E_p = E + p\hbar\omega$.

The term ${\bf P}_{\alpha\beta}^{(sh)}$ contributes both at finite
as well as at zero temperature. However it vanishes if the pump
does not work. This is so, since for a stationary scatterer only
energy conserving transitions are allowed. In Eq. (\ref{Eq10C}),
for the stationary scatterer, only the terms with $n=m=p=0$
remain. For such terms the difference $f_0(E_n) - f_0(E_m)$ is
identically zero.

Using the unitarity conditions Eqs. (\ref{Eq4}) we can check that
the above noise power is subject to the conservation laws:
\cite{Buttiker92}
\begin{equation}
\label{Eq11} \sum\limits_{\beta}{\bf P}_{\alpha\beta} = 0, \quad
\sum\limits_{\alpha}{\bf P}_{\alpha\beta} = 0.
\end{equation}
\noindent To prove the second equality one needs to make the
shifts $E\to E - p\hbar\omega$, $n\to n-p$, and $m\to m-p$ in
Eq. (\ref{Eq10C}).

Let us consider the sign of current correlations. The auto
correlations ($\alpha = \beta$) are positive and the cross
correlations ($\alpha \neq \beta$) are negative as it should be
for fermions. \cite{Buttiker92,BB00} The thermal noise ${\bf
P}_{\alpha\beta}^{(th)}(E)$ satisfies obviously this sign rule. To
show this for the shot noise contribution we follow
Ref.~\onlinecite{Buttiker92} and rewrite Eq. (\ref{Eq10C}):
\begin{widetext}
\begin{equation}
\label{Eq12}
{\bf P}_{\alpha\beta}^{(sh)}(E) = - \sum\limits_{p=-\infty}^{\infty}
\bigg| \sum\limits_{n=-\infty}^{\infty}
\sum\limits_{\gamma} f_0(E_n)
S_{F,\alpha\gamma}^{*}(E,E_n) S_{F,\beta\gamma}(E_p,E_n) 
\bigg|^2 < 0,
\quad \alpha\neq\beta.
\end{equation}
\end{widetext}
\noindent Since the cross correlations are negative and because of
the sum rule Eq. (\ref{Eq11}) we conclude that the auto
correlations are necessarily positive ${\bf
P}_{\alpha\alpha}^{(sh)}(E) > 0$.

We remark that cross-correlations can be positive 
even for normal conductors if they are imbedded in an external 
circuit with a finite impedance, and in particular even if 
they are subject to oscillating voltages applied to the
terminals of the conductor. In addition interactions can change 
the sign of correlations. 
We  refer the interested reader to recent discussions \cite{mb,cottet}. 
However, none of these factors plays a role 
here and the cross-correlations are negative as in a sample 
subject to dc-transport and connected to an ideal zero-impedance 
external circuit. 

Strictly speaking one can not rigorously separate the shot noise
from the thermal noise especially at $\hbar\omega\sim k_BT$. Here
we have introduced the two contributions, ${\bf
P}_{\alpha\beta}^{(th)}$ and ${\bf P}_{\alpha\beta}^{(sh)}$, for
convenience only. Their distinction is insofar justified as they
depend differently on the temperature and pump frequency.

\subsection{Heat flow fluctuations}
\label{hff}

By analogy with the current operator $\hat I_{\alpha}$ Eq. (\ref{Eq8})
we introduce a heat flow operator in the lead $\alpha$ :
\begin{equation}
\label{Eq13}
\begin{array}{c}
\hat I_{E,\alpha}(t) = \frac{1}{h} \int dEdE^{\prime} (E-\mu) \\
\ \\
\times
[\hat b^{\dagger}_{\alpha}(E)\hat b_{\alpha}(E^{\prime})
- \hat a^{\dagger}_{\alpha}(E)\hat a_{\alpha}(E^{\prime})]
e^{i\frac{E-E^{\prime}}{\hbar}t}.
\end{array}
\end{equation}

The heat flow $\hat I_{E,\alpha}$ is the difference between the
flows of energy (measured with respect to the Fermi energy $\mu$)
carried by electrons from the scatterer to the reservoir $\alpha$
and in the opposite direction. This definition makes sense since
the operators ($\hat b^{\dagger}_{\alpha}(E), \hat
a^{\dagger}_{\alpha}(E), \dots$) correspond to particles with a
definite energy $E$.

The correlation function ${\bf P}_{E,\alpha\beta}(t_{1}, t_{2})$
of heat flows and the zero frequency heat flow noise power ${\bf
P}_{E,\alpha\beta}$ of interest here is defined in a full analogy
with  Eqs. (\ref{Eq6}) and (\ref{Eq9}), [replacing $\hat
I_{\alpha}(t)$ by $\hat I_{E,\alpha}(t)$].

Using Eq. (\ref{Eq5}) we express the zero frequency heat flow noise
power ${\bf P}_{E,\alpha\beta}$ in terms of the Floquet scattering
matrix as follows:
\begin{widetext}
\begin{subequations}
\label{Eq14}
\begin{equation}
\label{Eq14A}
{\bf P}_{E,\alpha\beta} = \frac{2}{h}\int\limits_{0}^{\infty}dE
\Big( {\bf P}_{E,\alpha\beta}^{(th)}(E) + 
{\bf P}_{E,\alpha\beta}^{(sh)}(E) \Big) ,
\end{equation}
\begin{equation}
\label{Eq14B}
\begin{array}{c}
{\bf P}_{E,\alpha\beta}^{(th)}(E) = k_BT \left( - \frac{\partial 
f_0(E)}{\partial E} \right)
\Bigg( \delta_{\alpha\beta}(E-\mu)^2 +
\sum\limits_{n=-\infty}^{\infty}
\bigg\{ \delta_{\alpha\beta} (E_n - \mu)^2
\sum\limits_{\gamma} \big|S_{F,\alpha\gamma}(E_n,E)\big|^2  \\
\ \\
- (E-\mu)(E_n - \mu)
\Big[ \big|S_{F,\alpha\beta}(E_n,E)\big|^2  + 
\big|S_{F,\beta\alpha}(E_n,E)\big|^2 \Big]
\bigg\} \Bigg).
\end{array}
\end{equation}
\begin{equation}
\label{Eq14C}
{\bf P}_{E,\alpha\beta}^{(sh)}(E) = \sum\limits_{\gamma,\delta}
\sum\limits_{n=-\infty}^{\infty} 
\sum\limits_{m=-\infty}^{\infty} \sum\limits_{p=-\infty}^{\infty}
(E-\mu)(E_p - \mu)
\frac{[f_0(E_n) - f_0(E_m) ]^2}{2}
S_{F,\alpha\gamma}^{*}(E,E_n) S_{F,\alpha\delta}(E,E_m)
S_{F,\beta\delta}^{*}(E_p,E_m) S_{F,\beta\gamma}(E_p,E_n).
\end{equation}
\end{subequations}
\end{widetext}

In analogy with the current noise power we divide the whole heat
flow noise power into two parts. ${\bf P}_{E,\alpha\beta}^{(th)}$
is termed the thermal heat noise since it disappears at zero
temperature. ${\bf P}_{E,\alpha\beta}^{(sh)}(E)$ is called the
heat flow shot noise.

Since the pump is a source of energy flows from the scatterer to
the reservoirs for the heat flow noise power ${\bf
P}_{E,\alpha\beta}$ we do not expect a conservation law like
Eq. (\ref{Eq11}).

\subsection{Adiabatic approximation}
\label{AA}

To calculate the noise power ${\bf P}_{\alpha\beta}$ one needs to
know the Floquet scattering matrix. That requires the solution of
a full time-dependent scattering problem. However for a slowly
(adiabatically) oscillating scatterer the solution of a stationary
scattering problem is sufficient. \cite{MBstrong02,MBac03}

Let us assume that the scattering of electrons with energy $E$ by
the stationary sample can be described via the (stationary)
scattering matrix $\hat S(E,\{x\})$ depending on some parameters
$x_i \in \{X\} , i = 1,2,\dots ,N_p$ (e.g., sample's shape, the
strength of coupling to leads, etc.). Varying these parameters one
can change the scattering properties of a sample. We suppose these
parameters to be periodic functions in time: $x_{i}(t) =
x_{i}(t+{\cal T}) , \forall i$. Then the matrix $\hat S$ becomes
dependent on time: $\hat S(E,t) = \hat S(E,\{x(t)\})$. In general
the matrix $\hat S(E,t)$ does not describe the scattering of
electrons by the time-dependent scatterer: only the Floquet
scattering matrix $\hat S_{F}(E_n,E)$ does.

However if the parameters $x_i$ change so slowly that the energy
quantum $\hbar\omega$ is small compared with the relevant energy
scale $\delta E$ over which the scattering matrix $\hat S(E)$
changes significantly
\begin{equation}
\label{Eq15}
\hbar\omega \ll \delta E \ll \mu,
\end{equation}

\noindent and therefore $\hat S(E)\approx \hat S(E+\hbar\omega)$,
then the elements of the Floquet scattering matrix can be
expressed in terms of the Fourier coefficients of the matrix $\hat
S(t)$. Note the energy scale $\delta E$ can vary from case to
case. For instance, close to a transmission resonance $\delta E$
is a resonance width but far from the resonance $\delta E$ is
rather the distance between resonances. For many channel
scatterers well connected to leads $\delta E$ is the Thouless
energy.

The first two terms of an expansion of the Floquet 
scattering matrix in powers of $\hbar\omega/\delta
E \ll 1$ read \cite{MBac03}
\begin{equation}
\label{Eq16}
\hat S_{F}(E_n,E) = \hat S_{n}\left(\frac{E_{n}+E}{2}\right)
+ \hbar\omega\hat A_{n}(E) + O\big([\hbar\omega/\delta E]^2\big) .
\end{equation}

\noindent Here the matrix of Fourier coefficients is defined in
the following way ($Y \equiv A,S$):
\begin{equation}
\label{Eq17}
\begin{array}{c}
\hat Y_{n}(E) = \frac{\omega}{2\pi}\int\limits_{0}^{\cal T} dt 
e^{in\omega t} \hat Y(E,t), \\
\ \\
\hat Y(E,t) = \sum\limits_{n=-\infty}^{\infty}e^{-in\omega t} \hat
Y_{n}(E).
\end{array}
\end{equation}

The matrix $\hat A$ is subject to the following equation
\begin{subequations}
\label{Eq18}
\begin{equation}
\label{Eq18A}
\hbar\omega\left(\hat S^{\dagger}(E,t)\hat A(E,t) 
+ \hat A^{\dagger}(E,t)\hat S(E,t)\right)
= \frac{1}{2}{\cal P}\{\hat S^{\dagger};\hat S \},
\end{equation}
\begin{equation}
\label{Eq18B}
{\cal P}\{\hat S^{\dagger};\hat S \} =
i\hbar \left( \frac{\partial \hat S^{\dagger}}{\partial t}
\frac{\partial \hat S}{\partial E} -
\frac{\partial \hat S^{\dagger}}{\partial E}
\frac{\partial \hat S}{\partial t}
\right).
\end{equation}
\end{subequations}
\noindent Note the matrix ${\cal P}\{\hat S^{\dagger};\hat S \}$
is traceless.

Without magnetic fields when the stationary scattering matrix is
symmetric in lead indices, $S_{\alpha\beta} = S_{\beta\alpha}$,
the matrix $\hat A$ is antisymmetric: \cite{MBac03}
$A_{\alpha\beta}  = - A_{\beta\alpha}$.

\section{Adiabatic current noise power}
\label{ACNP}

In this section we use the adiabatic approximation 
Eqs. (\ref{Eq16}), (\ref{Eq18})
to calculate the current noise power Eq. (\ref{Eq10}).
A working pump modifies the thermal noise according to Eq. (\ref{Eq10B})
and it is a source of shot noise Eq. (\ref{Eq10C}).

\subsection{Current thermal noise}

In the high temperature regime, relevant for existing experiments,
\begin{equation}
\label{Eq19}
\hbar\omega \ll k_BT,
\end{equation}
\noindent the main source of noise are equilibrium thermal
fluctuations. A working pump modifies the thermal noise only
slightly. However the noise produced by the pump has a
characteristic dependence on frequency and temperature and thus
can be detected.

To calculate the thermal noise ${\bf P}^{(th)}_{\alpha\beta}$ in
the presence of a slowly oscillating scatterer we substitute the
adiabatic expansion Eqs. (\ref{Eq16}) for the Floquet scattering
matrix into Eq. (\ref{Eq10B}). Since we find the Floquet scattering
matrix with an accuracy of $\omega$ we calculate the noise with the
same accuracy.

After a little algebra (see Appendix, Sec.\ref{A0}) we find
for the zero-frequency thermal noise:
\begin{subequations}
\label{Eq20}
\begin{equation}
\label{Eq20A}
{\bf P}^{(th)}_{\alpha\beta} = 
\frac{2e^2}{h}\int\limits_{0}^{\infty}dE
\Big( {\bf P}^{(th,0)}_{\alpha\beta}(E) 
+ {\bf P}^{(th,p)}_{\alpha\beta}(E) \Big),
\end{equation}
\begin{equation}
\label{Eq20B}
{\bf P}^{(th,0)}_{\alpha\beta}(E) =
 k_BT \left( -\frac{\partial f_0}{\partial E}\right)
\left(2\delta_{\alpha\beta} - \overline{\big|S_{\alpha\beta}(E) \big|^2}
- \overline{\big|S_{\beta\alpha}(E) \big|^2}
\right),
\end{equation}
\begin{equation}
\label{Eq20C}
{\bf P}^{(th,p)}_{\alpha\beta}(E) = 
k_BT \left( -\frac{\partial f_0}{\partial E}\right)\frac{h}{e} \left(
\delta_{\alpha\beta} \frac{dI_{\alpha}(E)}{dE} 
- \frac{dI_{\alpha\beta}^{(s)}(E)}{dE}\right).
\end{equation}
\end{subequations}

The first term ${\bf P}^{(th,0)}_{\alpha\beta}$ is the equilibrium
(Nyquist-Johnson) noise in the presence of a working pump. The
second term ${\bf P}^{(th,p)}_{\alpha\beta}$ is a contribution to
the thermal noise due to the operating pump. For an adiabatic pump
the first term is thus independent of frequency whereas the second
term is linear in the pump frequency $\omega$ [see,
Eqs. (\ref{Eq21}), (\ref{Eq22})].

In Eq. (\ref{Eq20B}) we have introduced the time averaged conductances
$$
\overline{\big|S_{\alpha\beta}(E) \big|^2} =
\int\limits_{0}^{\cal T} \frac{dt}{\cal T} 
\big|S_{\alpha\beta}(E,t) \big|^2.
$$

In Eq. (\ref{Eq20C}) we have introduced the quantities which are
intrinsic characteristics of a working pump: the instantaneous
currents generated by the pump. \cite{MBac03} The first one
$dI_{\alpha}/dE$ is the spectral density of (dc) currents pushed
by the pump into the lead $\alpha$:
\begin{equation}
\label{Eq21}
\frac{dI_{\alpha}(E)}{dE} = i\frac{e}{2\pi}  \int\limits_{0}^{\cal T} 
\frac{dt}{\cal T}\left(  \frac{\partial\hat S}{\partial t}
\frac{\partial\hat S^{\dagger}}{\partial E} 
-\frac{\partial\hat S}{\partial E}
\frac{\partial\hat S^{\dagger}}{\partial t}\right)_{\alpha\alpha}.
\end{equation}

The second quantity is a symmetrized spectral current density
describing currents driven from lead $\alpha$ to lead $\beta$:
$dI_{\alpha\beta}^{(s)}/dE = dI_{\alpha\beta}/dE +
dI_{\beta\alpha}/dE$, where  $dI_{\alpha\beta}/dE$ is the spectral
current density pushed by the pump from lead $\beta$ to lead $\alpha$. 
Without magnetic fields the scattering matrix $\hat S$ is symmetric in the lead indices
and the matrix $\hat A$ is antisymmetric.
In this case we find: \cite{MBac03}
\begin{equation}
\label{Eq22}
\frac{dI_{\alpha\beta}^{(s)}(E)}{dE} = 
i\frac{e}{2\pi}  \int\limits_{0}^{\cal T} \frac{dt}{\cal T}
\left(  \frac{\partial S_{\alpha\beta}}{\partial t}
\frac{\partial S_{\alpha\beta}^{*}}{\partial E} -
\frac{\partial S_{\alpha\beta}}{\partial E}
\frac{\partial S_{\alpha\beta}^{*}}{\partial t}
\right).
\end{equation}

Clearly, we have $\sum\limits_{\beta}dI_{\alpha\beta}^{(s)}/dE =
dI_{\alpha}/dE$. Therefore each of the terms ${\bf
P}^{(th,0)}_{\alpha\beta}$ and ${\bf P}^{(th,p)}_{\alpha\beta}$
separately satisfies the conservation laws Eq. (\ref{Eq11}).

The contribution ${\bf P}^{(th,p)}_{\alpha\beta}$ is the result of
an interference between fluctuating currents coming from the
reservoirs and ac currents generated by the pump. Interference
terms of this type occur also for a pump in the presence of
oscillating reservoir potentials. \cite{MB03} Such interference
terms are thus present whenever the reservoirs are brought into a
non-equilibrium state either through ac-voltages or as here through
spontaneous equilibrium fluctuations.
These interference corrections are small compared to the
equilibrium thermal noise,
\begin{equation}
\label{Eq23} \frac{{\bf P}^{(th,p)}_{\alpha\beta}}{{\bf
P}^{(th,0)}_{\alpha\beta}} \sim \frac{\hbar\omega}{\delta E}.
\end{equation}
\noindent Nevertheless it is these interference corrections 
which are responsible for the dependence
of the high temperature current noise, Eq. (\ref{Eq29}), on the pump
frequency $\omega$.

The pump induced contribution to the thermal noise ${\bf
P}^{(th,p)}_{\alpha\beta}$ is linear in frequency $\omega$ like
the pumped current. On the other hand it is linear in temperature
$T$ like the equilibrium noise itself. However unlike the ${\bf
P}^{(th,0)}_{\alpha\beta}$, which is positive at $\alpha=\beta$
and negative at $\alpha\neq\beta$, the interference corrections
${\bf P}^{(th,p)}_{\alpha\beta}$ being small have no definite
sign. The instantaneous currents $dI_{\alpha\beta}/dE$ produced by
the pump can be either positive (flowing from lead $\beta$ to 
lead $\alpha$) or negative (flowing from lead $\alpha$ to
lead $\beta$). For instance, in Sec.\ref{SE} we consider a
particular example where the pump produces a negative contribution
to the auto correlations ($\alpha = \beta$) and a positive correction
to the cross correlations ($\alpha\neq\beta$). We emphasize that this
fact does not change the sign of the whole thermal noise ${\bf
P}_{\alpha\beta} = {\bf P}^{(th,0)}_{\alpha\beta} + {\bf
P}^{(th,p)}_{\alpha\beta}$ which satisfies the sign rules for
fermionic systems. \cite{BB00} This follows from the general
equation (\ref{Eq10B}).

\subsection{Current shot noise}

\subsubsection{Zero temperature}
\label{ANP_ZT}

At zero temperature $T=0$, in the zero-frequency limit considered
here, only  a working pump is a source of a noise. The noise
arises because the pump generates a non-thermal distribution of
outgoing particles with a characteristic energy scale of order
$\hbar\omega$. This noise is quite analogous (but not identical)
to the shot noise arising in dc-biased (with $eV \sim
\hbar\omega$) conductors. \cite{BB00,MB02}

Substituting the adiabatic expansion for the Floquet scattering
matrix Eq. (\ref{Eq16}) in Eq. (\ref{Eq10}) and keeping only leading
in $\omega$ terms we get the zero temperature noise power as
follows:
\begin{subequations}
\label{Eq24}
\begin{equation}
\label{Eq24A}
{\bf P}_{\alpha\beta} = \frac{2e^2}{h}
\sum\limits_{q=1}^{\infty} q\hbar\omega 
C_{\alpha\beta,q}^{(sym)}(\mu),
\end{equation}
\begin{equation}
\label{Eq24B}
C_{\alpha\beta,q}^{(sym)}(E) = 
\frac{C_{\alpha\beta,q}(E) + C_{\alpha\beta,-q}(E) }{2},
\end{equation}
\begin{equation}
\label{Eq24C}
C_{\alpha\beta,q}(E) =
\sum\limits_{\gamma} \sum\limits_{\delta}
\big(S_{\alpha\gamma}^{*}(E) S_{\alpha\delta}(E) \big)_{q}
\big(S_{\beta\delta}^{*}(E) S_{\beta\gamma}(E) \big)_{-q}.
\end{equation}
\end{subequations}
\noindent Here we have introduced the matrix $\hat C_{q}$ which we
relate to a two-particle scattering matrix in Appendix,
Sec.\ref{A1}. The factor $q\hbar\omega$, with $q = |n-m|$, is the
size of an energy window for electrons where the conditions for
quantum mechanical exchange (see below) which is responsible for
the shot noise are fulfilled.

In Eq. (\ref{Eq24A}) the matrix elements
$C_{\alpha\beta,q}^{(sym)}(\mu)$ are evaluated at the Fermi
energy: $E=\mu$. Integrating over energy we take into account the
fact that the adiabatic scattering matrix by definition [see,
Eq. (\ref{Eq15})] is independent of energy over the scale of order
$\hbar\omega$. The lower index $\pm q$ denotes the Fourier
coefficients.

To clarify the structure of the expression given above, let us
consider the scattering process relevant for the zero frequency
noise. The basic process consists of scattering of two (non
correlated) particles. \cite{BB00} These particles with energies
$E_n = E + n\hbar\omega$ and $E_m = E + m\hbar\omega$ enter the
sample through the leads $\gamma$ and  $\delta$. Due to the
interaction with the time-dependent scatterer they emit/absorb
some energy quanta $\hbar\omega$ and leave the scatterer with
energies $E$ and $E_p$ through the leads $\alpha$ and $\beta$.

The origin of noise at zero temperature is a quantum mechanical
exchange. This effect matters for those outgoing energies $E$ for
which simultaneously one incoming state, say corresponding to the
energy $E_n$ in lead $\gamma$, is full (there is an incident
particle) and another incoming state, corresponding to the energy
$E_m$ in lead $\delta$, is empty (there is an incident "hole") or
vice versa. Note, in general,  the incoming states have different
energies $E_n\neq E_m$. Therefore these states can be subject to
the exchange effect only if the energy difference $q\hbar\omega =
|E_n - E_m|$ will be compensated by the oscillating scatterer.
Hence we are dealing with {\it photon-assisted exchange}. 
As a consequence of quantum mechanical exchange, the particle can
appear with energy $E$ in lead $\alpha$ or with energy $E_p$ in
lead $\beta$.

Note that Eq. (\ref{Eq24}) is only the leading term of the series
expansion of the noise in powers of $\omega$. In principle the
approximation Eq. (\ref{Eq16}) is sufficient to calculate the next
term of the zero temperature noise power which is of order
$\hbar\omega^2/\delta E$.

\subsubsection{High temperature: $k_BT\gg\hbar\omega$}
\label{ANP_HT}

At high temperature $k_BT\gg\hbar\omega$ the shot noise is only a
part of the entire noise. In this case the term ${\bf
P}_{\alpha\beta}^{(sh)}$ Eq. (\ref{Eq10C}) gives (in leading order)
rise to a term proportional to $\sim \omega^2$ in the current
correlations. Strictly speaking we are unable to calculate the
entire noise with an accuracy of order $\omega^2$. To this end one
needs to know the Floquet scattering matrix with the same
accuracy. This is beyond the approximation Eq. (\ref{Eq16}) used
here. 
In particular, the term ${\bf P}^{(th)}_{\alpha\beta}$,
Eq. (\ref{Eq10B}), would lead to noise of the order $\sim
k_BT\big(\hbar\omega/\delta E\big)^2$.
However the shot noise Eq. (\ref{Eq10C}) has a characteristic
temperature dependence (it is proportional to the inverse temperature) and
thus can be distinguished in experiment from the thermal ($\sim
T$) noise.

Calculating Eq. (\ref{Eq10C}) to leading order in $\omega$ we use
the lowest adiabatic approximation for the Floquet scattering
matrix: $\hat S_{F}(E_n,E_m) \approx \hat S_{n-m}(E)$,  and find
\begin{equation}
\label{Eq25} {\bf P}_{\alpha\beta}^{(sh)} =
\frac{2e^2}{h}\int\limits_{0}^{\infty}dE \left( \frac{\partial
f_0}{\partial E}\right)^2 \sum\limits_{q=1}^{\infty}
(q\hbar\omega)^2 C_{\alpha\beta,q}^{(sym)}(E).
\end{equation}

This quadratic in $\omega$ noise power is (for $\alpha=\beta$)
exactly what was calculated by Avron et al. \cite{AEGS01}.

\subsubsection{Weakly energy dependent scattering matrix}
\label{EISM}

If the temperature $T$ is small compared to the relevant energy
scale $\delta E$ for the scattering matrix
\begin{equation}
\label{Eq26}
k_BT \ll \delta E,
\end{equation}
then integrating over energy in Eq. (\ref{Eq10}) we can keep the
scattering matrix as energy independent. 
As a result we obtain expression for the shot noise valid 
at arbitrary ratio $\hbar\omega/k_BT$.

Since the shot noise itself is proportional to $\omega$ then to
calculate it in leading order in $\omega$ we can use the Floquet
scattering matrix in the lowest adiabatic approximation. Thus the
leading order shot noise is:
\begin{subequations}
\label{Eq27}
\begin{equation}
\label{Eq27A} {\bf P}_{\alpha\beta}^{(sh)} = \frac{2e^2}{h}
\sum\limits_{q=1}^{\infty}F(q\hbar\omega,k_BT)
C_{\alpha\beta,q}^{(sym)}(\mu),
\end{equation}
\begin{equation}
\label{Eq27B}
F(q\hbar\omega,k_BT) = 
q\hbar\omega \coth\left(\frac{q\hbar\omega}{2k_BT} \right) - 2k_BT,
\end{equation}
\end{subequations}

The first part on the RHS of Eq. (\ref{Eq27B}) is well known (with
$q=1$) from the theory of frequency dependent noise \cite{BB00}
and from the fluctuation-dissipation theorem [see, e.g.,
Ref.~\onlinecite{LL5}]. The second part on the RHS of
Eq. (\ref{Eq27B}) (i.e., $-2k_BT$) just compensates the high
temperature contribution included already into the thermal noise
Eqs. (\ref{Eq10B}) and (\ref{Eq20B}).

The equation (\ref{Eq27A}) reproduces both the equation
(\ref{Eq24A}) at low temperatures $k_BT\ll\hbar\omega$ and the
equation (\ref{Eq25}) at high temperatures $k_BT\gg\hbar\omega$.

The shot noise, Eq. (\ref{Eq27}), satisfies the conservation laws
Eq. (\ref{Eq11}). That follows directly from the sum rule for the
matrix $\hat C(\tau)$, Eq. (\ref{EqA_7}).

The sign rule for the shot noise follows from Eq. (\ref{Eq12}).
However one can verify it directly in Eq. (\ref{Eq27A}). To this
end we perform the inverse Fourier transformation and represent
the shot noise, Eq. (\ref{Eq27A}), in terms of a time integral:
\begin{equation}
\label{Eq28}
\begin{array}{c}
{\bf P}_{\alpha\beta}^{(sh)} = \frac{2e^2}{h} \hbar\omega
\int\limits_{0}^{{\cal T}} \frac{d\tau}{{\cal T}} K(\tau) 
\Big( \delta_{\alpha\beta} - C_{\alpha\beta}^{(sym)}(\mu,\tau) \Big), \\
\ \\
K(\tau) = \left(\frac{\pi k_BT}{\hbar\omega}\right)^2
\sum\limits_{k=-\infty}^{\infty}
\sinh^{-2}\left(\frac{\pi k_BT}{\hbar\omega}\big[2\pi k 
+ \omega \tau \big]  \right).
\end{array}
\end{equation}

Since $C_{\alpha\beta}(\tau)$ is positively defined [see, Eq. (\ref{EqA_6})]
and it is less then unity [see, Eq. (\ref{EqA_7})],
the shot noise ${\bf P}_{\alpha\beta}^{(sh)}$ 
is positive at $\alpha=\beta$ and negative at $\alpha\neq\beta$.

At zero temperature $T=0$, we have $K(\tau) =
\frac{1}{2}[1-\cos(\omega \tau)]^{-1}$, and for $\alpha=\beta$ the
equation (\ref{Eq28}) agrees with that obtained within the
framework of the theory of full counting statistics.
\cite{AK00,MM01}

\subsection{Frequency dependence of the pump current noise}
\label{FDPN}

At zero temperature the pump produces only shot noise,
Eq. (\ref{Eq24A}), which is linear in pump frequency $\omega$. 
At nonzero temperatures the pump modifies the thermal noise 
and produces corrections to it which depend on pump frequency.
Therefore the full $\omega$-dependent  noise 
$\delta {\bf P}_{\alpha\beta}$ due to the
working pump is a sum of two contributions: $\delta {\bf
P}_{\alpha\beta} = {\bf P}^{(th,p)}_{\alpha\beta} +  {\bf
P}_{\alpha\beta}^{(sh)}$. The first one is a correction to the
thermal noise, Eq. (\ref{Eq20C}), (integrated over energy). This
contribution is linear in pump frequency. The second one is a high
temperature shot noise Eq. (\ref{Eq25})  which is quadratic in pump
frequency. Let us compare them. The additional thermal noise
produced by the adiabatic ($\hbar\omega \ll \delta E$) pump is of
order ${\bf P}^{(th,p)}_{\alpha\beta} \sim
k_BT\frac{\hbar\omega}{\delta E}$, where $\delta E$ is the energy
scale of $\hat S(E)$. The high temperature shot noise is of the
order of ${\bf P}_{\alpha\beta}^{(sh)} \sim
\frac{(\hbar\omega)^2}{k_BT}$. 
Their ratio is ${\bf P}^{(sh)}_{\alpha\beta}/{\bf P}_{\alpha\beta}^{(th,p)} \sim
\frac{\hbar\omega\delta E}{(k_BT)^2}$. 
At relatively high temperatures, $k_BT\geq\delta E$, it is 
${\bf P}^{(sh)}_{\alpha\beta}/{\bf P}_{\alpha\beta}^{(th,p)} \sim
\frac{\hbar\omega}{k_BT}$.

Therefore at lower
temperatures the shot noise ($\sim \omega^2$) still prevails
whereas at higher temperatures the pump produces mainly a linear
in $\omega$ thermal-like noise.

Thus we expect that with increasing temperature 
the dependence on the pump frequency 
of the experimentally detectable additional noise power
changes from linear to quadratic and again to linear:
\begin{equation}
\label{Eq29}
\delta {\bf P}_{\alpha\beta} \sim \left\{
\begin{array}{ll}
\omega, &  k_BT \ll \hbar\omega, \\
\ \\
\omega^2, & \hbar\omega \ll k_BT \ll \sqrt{\hbar\omega \delta E}, \\
\ \\
\omega, & \sqrt{\hbar\omega \delta E} \ll k_BT.
\end{array}
\right.
\end{equation}

We reemphasize that different physical mechanisms are responsible
for the linear in $\omega$ behavior of the noise power at zero and
at high temperatures. In the former case it is a shot noise
whereas in the latter case it is a thermal-like noise.

\section{Adiabatic heat flow fluctuations: 
            weakly energy dependent scattering matrix}
\label{AHFF}

Like in the previous section we use the adiabatic approximation
given by 
Eqs. (\ref{Eq16}) and (\ref{Eq18}). We calculate the heat flow noise
power Eq. (\ref{Eq14}) at relatively low temperatures:
$k_BT\ll\delta E$. In this case, when integrating
over energy in Eq. (\ref{Eq14}), we can treat the scattering
matrix as energy independent and take it at the Fermi energy
$\mu$. We calculate the noise to leading order in the pump
frequency $\omega$.

\subsection{Heat thermal noise}

If the temperature exceeds the pump frequency
$k_BT\gg\hbar\omega$, Eq. (\ref{Eq19}), then the thermal noise
${\bf P}_{E,\alpha\beta}^{(th)}$ dominates over the shot noise
${\bf P}_{E,\alpha\beta}^{(sh)}$. The heat thermal noise consists
of two parts: an equilibrium Nyquist - Johnson heat flow noise
\cite{KBSL01} and a linear in $\omega$ correction due to an
operating pump. The last term is similar to a thermal current
noise, Eq. (\ref{Eq20}).

Keeping $\hat S(E)\approx {\rm const}$ we can relate the heat thermal 
noise  ${\bf P}_{E,\alpha\beta}^{(th)}$, Eq. (\ref{Eq14B}), 
(integrated over energy) to the current thermal noise 
${\bf P}_{\alpha\beta}^{(th)}$, Eq. (\ref{Eq20A}) in a very simple way
\begin{equation}
\label{Eq30} \frac{{\bf P}_{E,\alpha\beta}^{(th)}}{{\bf
P}_{\alpha\beta}^{(th)}} = \frac{\pi^2}{3e^2}(k_BT)^2, \quad k_BT
\ll \delta E.
\end{equation}

Such a ratio is quite expected for the equilibrium thermal noises.
That follows from the fluctuation-dissipation theorem (FDT) [see,
e.g., Ref.~\onlinecite{LL5}] relating the equilibrium fluctuations
and the corresponding linear response functions. According to the
FDT the current and heat fluctuations in the equilibrium state are
proportional to the (linear response) electrical $\sigma$ and
thermal $\kappa$ conductivity, respectively. If electrons are
subject to only elastic scattering then the above conductivities
are related through the Wiedemann-Franz law:
$\frac{\kappa}{\sigma} = \frac{\pi^2}{3e^2}(k_BT)^2$ [see, e.g.,
Ref.~\onlinecite{A87}]. For the mesoscopic (phase coherent)
samples with energy independent scattering matrix the
Wiedemann-Franz law (with the same ratio) can be formulated in
terms of electrical $G$ and thermal ${\cal K}$ conductances which
characterize the entire sample. \cite{EA81,SI86,M98} As a result we
arrive at equation (\ref{Eq30}).

It should be noted that the equation (\ref{Eq30}) holds for the
thermal noise modified by the pump. Therefore we can suggest that
the instantaneous currents $dI_{\alpha}/dE$ Eq. (\ref{Eq21})
generated by the pump can be viewed as quasi-equilibrium  currents
(at least at high temperature).

\subsection{Heat flow shot noise}

\subsubsection{Low temperature: $k_BT\ll\hbar\omega$}

When the temperature is lowered below the energy of the modulation
quanta $\hbar\omega$ the shot noise ${\bf
P}_{E,\alpha\beta}^{(sh)}$ becomes dominant. This noise is more
subtle and can not be related in a simple way to the current shot
noise ${\bf P}_{\alpha\beta}^{(sh)}$, Eq. (\ref{Eq10C}), even for
an energy independent scattering matrix. The reason is that unlike
the current shot noise the heat flow shot noise, Eq. (\ref{Eq14C}),
is sensitive to the actual energies of outgoing particles.

To clarify this difference we integrate over energy in Eq. (\ref{Eq14C}). 
We can do this 
since the adiabatic scattering
matrix is independent of energy over the interval relevant for
such an integration at $k_BT\ll \hbar\omega$. 
Introducing the difference between the
energies of incoming states $|E_n - E_m| = q\hbar\omega$ we find:
\begin{equation}
\label{Eq31}
\begin{array}{c}
\int\limits_{0}^{\infty}dE(E-\mu)(E_p-\mu)
\frac{[f_0(E_n)-f_0(E_m)]^2}{2} =  \\
\ \\
q\hbar\omega \left( -\frac{(E_n - E_m)^2}{12}
+ \frac{(E - E_n)(E_p - E_n) + (E - E_m)(E_p - E_m)} {4} \right).
\end{array}
\end{equation}

The first term (equal to $-(q\hbar\omega)^2/12$) in the big round
brackets on the RHS of Eq. (\ref{Eq31}) depends only on one energy
difference. Therefore the corresponding part of the heat flow
noise can be represented as a sum over $q$ by analogy with the
current shot noise, Eq. (\ref{Eq24}). In contrast, the second term
on the RHS of Eq. (\ref{Eq31}) depends on two energy differences
and thus it results in fluctuations of a completely different
nature.

After these preliminary remarks let us proceed with the calculation of
the heat flow shot noise, Eq. (\ref{Eq14C}). To obtain the noise to
leading order in the pump frequency $\omega$ we use the lowest
order adiabatic approximation for the Floquet scattering matrix
elements (e.g., $S_{F,\alpha\beta}(E_p,E_n)\approx
S_{\alpha\beta,p-n}$, etc.). Taking into account Eq. (\ref{Eq31})
and applying the inverse Fourier transformation (for all the
indices except the last one) we obtain:
\begin{equation}
\label{Eq32}
{\bf P}_{E,\alpha\beta}^{(sh)} = 
\frac{2}{h}\sum\limits_{q=1}^{\infty}q\hbar\omega \bigg(
- \frac{(q\hbar\omega)^2}{6}C^{(sym)}_{\alpha\beta,q}(\mu)
+   D^{(sym)}_{\alpha\beta,q}(\mu)
\bigg).
\end{equation}

As in the expression for the current shot noise Eq. (\ref{Eq24}),
the heat current shot noise contains the factor $q\hbar\omega$ 
which gives the size of the energy window for electrons where the
appropriate conditions for quantum mechanical exchange are fulfilled.

Equation (\ref{Eq32}) for the heat flow shot noise consists of two
parts related directly to the two terms of Eq. (\ref{Eq31}). The
first one is proportional to the matrix element
$C^{(sym)}_{\alpha\beta,q}$ entering the expression for the
current shot noise, Eq. (\ref{Eq24A}). The second part of
Eq. (\ref{Eq32}) is determined by the new matrix $\hat
D^{(sym)}_{q}$. The symmetrized elements are (see also, Appendix,
Sec.\ref{A2}):
\begin{equation}
\label{Eq33}
\begin{array}{c}
D_{\alpha\beta,q}^{(sym)} = 
\frac{D_{\alpha\beta,q} + D_{\alpha\beta,-q} }{2}, \\
\ \\
D_{\alpha\beta,q} = 
\frac{\hbar^2}{2} \sum\limits_{\gamma} \sum\limits_{\delta}
\left\{\left( S_{\alpha\gamma}^{*} 
\frac{\partial  S_{\alpha\delta} }{\partial t}  \right)_{q}
\left( S_{\beta\gamma}^{*} 
\frac{\partial  S_{\beta\delta} }{\partial t}  \right)_{q}^{*} \right. \\
\ \\
\left.
+ \left( \frac{\partial S_{\alpha\gamma}^{*} }{\partial t}  
S_{\alpha\delta}  \right)_{q}
   \left( \frac{\partial S_{\beta\gamma}^{*}   }{\partial t}  
S_{\beta\delta}    \right)_{q}^{*}
\right\}.
\end{array}
\end{equation}

The matrix $\hat D^{(sym)}_{q}$ differs essentially from the
matrix $\hat C^{(sym)}_{q}$. If the latter is defined by the
two-particle scattering matrix alone, Eq. (\ref{EqA_6}), the
former, in addition, depends on the energy shift matrix,
Eqs. (\ref{EqA_15}) - (\ref{EqA_17}). This allows us to consider
the second term in Eq. (\ref{Eq32}) as a manifestation of a novel
effect which is not visible in the current shot noise,
Eq. (\ref{Eq24A}).

Although both parts in Eq. (\ref{Eq32}) originate from quantum mechanical 
exchange the energy constraints are accounted for differently in the two terms. 
In the first part (proportional to $\hat C^{(sym)}_{q}$) 
quantum mechanical averaging is decoupled from averaging over energy 
(for the case when the scattering matrix can be taken to be energy independent). 
Therefore we can interpret it in the same way as the current noise, 
Eq. (\ref{Eq24}). 
In contrast, the second part (proportional to $\hat D^{(sym)}_{q}$) 
depends on the energy of both the incoming and out-going particles. 
These energy can be different owing to the pump compensating such difference. 
In this part averaging over energy affects essentially the result of
quantum mechanical averaging.

Concluding this section we briefly discuss the dependence of the
current-to-noise ratio on the pump frequency $\omega$. 
The pumped current and the zero temperature current noise both are linear in
the pump frequency. Therefore their ratio is independent of $\omega$. 
In contrast, the dc heat flow and its fluctuations depend on 
$\omega$ in different ways. 
While the zero temperature dc heat flow $I_{E,\alpha}$ 
is quadratic in pump frequency [see, e.g.,Refs.~\onlinecite{MB02,MBstrong02}], 
the zero temperature heat flow noise is proportional to $\omega^3$, 
Eq. (\ref{Eq32}). Therefore the ratio 
${\bf P}_{E,\alpha\alpha}/I_{E,\alpha}\sim \omega$ 
can be made arbitrary small in the adiabatic limit $\omega\to 0$.

\subsubsection{High temperature: $k_BT\gg\hbar\omega$}

With increasing temperature the difference between the current and
heat flow shot noises diminishes.

If the approximation of an energy independent scattering matrix,
Eq. (\ref{Eq26}), is appropriate, then the high temperature heat
flow shot noise is proportional to the current shot noise.
Integrating over energy in Eqs. (\ref{Eq14C}) and Eq. (\ref{Eq25})
we get
\begin{equation}
\label{Eq34} \frac{{\bf P}_{E,\alpha\beta}^{(sh)}}{{\bf
P}_{\alpha\beta}^{(sh)}} = \frac{3a}{e^2}(k_BT)^2, \quad
\hbar\omega\ll k_BT \ll \delta E.
\end{equation}
\noindent Here the factor is: $a =
\int\limits_{-\infty}^{\infty}dx x^2\cosh^{-4}(x) \approx 0.43$.
Note, the ratio in Eq. (\ref{Eq34}) differs from the ratio in
Eq. (\ref{Eq30}) only by the factor of $9a/(\pi)^2 \approx 0.39$.

\section{A simple example}
\label{SE}

In this section we illustrate the results obtained in sections
\ref{ACNP} and \ref{AHFF} performing numerical calculations for a
simple model.

Characterization of the noise generated by pumping is of interest
for all types of pumps but is of special interest in connection
with quantized charge pumping. Nearly quantized charge pumping can
be realized near resonant transmission peaks.
\cite{WWG00,LEWW01,EWA02,BDR03} 
Therefore it is instructive to investigate a model which exhibits 
a resonance-like transmission.

We consider a one-dimensional scatterer consisting of two delta
function barriers $V_{j}(x.t),~j = 1, 2$ oscillating with
frequency $\omega$ and located at $x=-L/2$ and $x=L/2$:
\cite{HN91}
\begin{equation}
\label{Eq35}
\begin{array}{l}
V_{1}(x,t) = (V_{01} 
+ 2V_{11}\cos(\omega t + \varphi_1))\delta(x+\frac{L}{2}), \\
\ \\
V_{2}(x,t) = (V_{02} 
+ 2V_{12}\cos(\omega t + \varphi_2))\delta(x-\frac{L}{2}), \\
\end{array}
\end{equation}
Thus $V_{01}$ and $V_{02}$ determine the static strength of the
barriers whereas $V_{11}$ and $V_{12}$ are the oscillatory
amplitudes which are used to pump charge. 

To calculate the noise power in the adiabatic limit 
it is enough to know the stationary scattering matrix $\hat S$.
Since the adiabatic condition $\hbar\omega \ll \delta E$
should hold over the whole pump cycle, 
which in our case crosses the resonance line, 
as a relevant energy scale $\delta E$ we choose
the smallest one characteristic for the scattering matrix, 
i.e., the width of a transmission resonance.

For the model under consideration the stationary scattering matrix is 
the following:
\begin{equation}
\label{Eq36}
 \hat S = \frac{e^{ikL}}{\Delta}
\left(
\begin{array}{cc}
\xi + 2\frac{p_2}{k}\sin(kL) & 1 \\
\ \\
1 & \xi + 2\frac{p_1}{k}\sin(kL)
\end{array}
\right).
\end{equation}

\noindent Here $k = \sqrt{\frac{2m}{\hbar^2}E}$; $p_j =
V_jm/\hbar^2$ (j = 1,2); $\xi$ = $(1-\Delta)e^{-ikL}$; $\Delta$ =
$1 + \frac{p_1p_2}{k^2}(e^{2ikL} -1)$ + $i\frac{p_1+p_2}{k}$. In
numerical calculations we use the units $2m = \hbar = e = 1$. 
We take $L = 100\pi$; $V_{01} = V_{02} = 20$ 
and investigate pumping for Fermi energies 
far exceeding an interlevel distance. We chose it close to the 101st
resonance state of the well which is at $k_F = 1.00968$.

\subsection{Current noise power}

In the two channels case ($\alpha,\beta = 1,2$) the current noise
power matrix $\hat {\bf P}$ can be entirely characterized by only
one matrix element. \cite{BB00} This is because of the
conservation law Eq. (\ref{Eq11}). To be definite we consider
current auto correlations ${\bf P}_{11}$.

 \begin{figure}[t]
  \vspace{3mm}
  \centerline{
   \epsfxsize8cm
 \epsffile{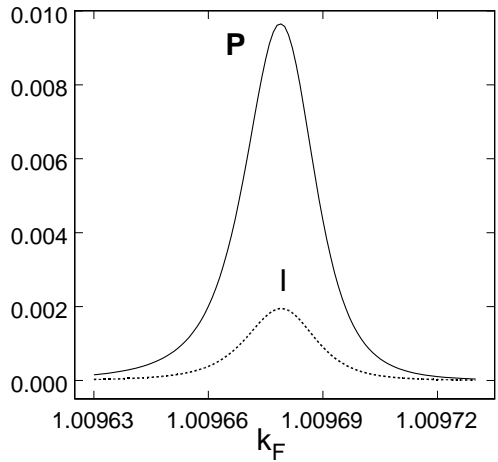}
             }
  \vspace{3mm}
  \nopagebreak
  \caption{
Current noise. Zero temperature weak pumping. The dependence of
the dc current $I \equiv I_{1}$, in units of $e\omega/(2\pi)$, and
the noise power  ${\bf P} \equiv {\bf P}_{11}$, in units of
$e^2\omega/\pi$, on the Fermi wave number $k_F$. The transmission
resonance through the double barrier structure is at $k_F =
1.00968$. The parameters are: $L = 100\pi$; $V_{01} = V_{02} =
20$; $V_{11} = V_{12} = 0.1$; $\varphi_2$ - $\varphi_1$ = $\pi
/2$.
   }
\label{fig1}
\end{figure}

\subsubsection{Zero temperature}
\label{SE_ZT}

\begin{figure}[t]
  \vspace{3mm}
  \centerline{
   \epsfxsize8cm
 \epsffile{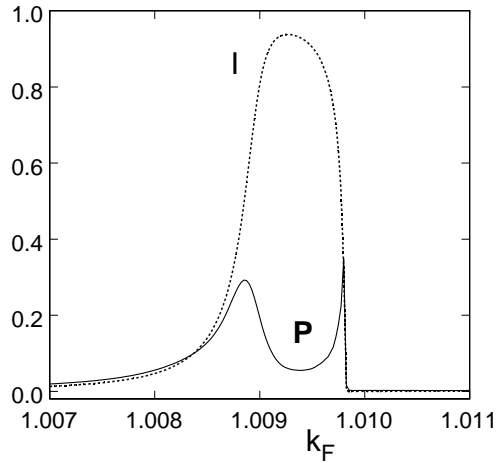}
             }
  \vspace{3mm}
  \nopagebreak
  \caption{
Current noise. Zero temperature strong pumping.
Parameters are the same as in Fig.\ref{fig1} except
that $V_{11} = V_{12} = 10$. Note the different scales
in comparison to Fig. {\ref{fig1}} .}
\label{fig2}
\end{figure}

At zero temperature only the shot noise matters. In Fig.\ref{fig1}
and Fig.\ref{fig2} we present the noise power ${\bf P}\equiv{\bf
P}_{11}$ together with the dc pumped current $I \equiv I_{1}$ as a
function of the Fermi wave number $k_F =
\sqrt{\frac{2m}{\hbar^2}\mu}$ in the vicinity of a transmission
resonance for weak ($V_{1j} \ll V_{0j}$) and strong ($V_{1j} \sim
V_{0j}$) pumping regimes, respectively. Note that at zero
temperature the pumped current $I_{\alpha}$ is: \cite{MBstrong02}
\begin{equation}
\label{Eq37}
I_{\alpha} = \omega \frac{e}{2\pi} \sum\limits_{\beta} 
\sum\limits_{q = -\infty}^{\infty}
q \left| S_{\alpha\beta, q}(\mu)\right|^2.
\end{equation}

We see that with increasing amplitude $V_{1j}$ of the oscillatory
pump parameters the magnitude of the noise power increases. 
At the same time relative to the pumped current the noise power
decreases. In addition, in the strong pumping regime the noise
shows a non monotonic behavior in the vicinity of the transmission
resonance. Note also that the scale of the vertical axis is very
different in Fig.\ref{fig1} and Fig.\ref{fig2}. While for a small
amplitude pump cycle we explore only the scattering properties in
the immediate vicinity of the resonance, the large amplitude pump
cycle can enclose the resonance even if the center of the pump
cycle is already "far" from resonance. For large amplitude pumping
the maximum pumped current is achieved if the resonance lies well
in the interior of the pump cycle.

 \begin{figure}[b]
  \vspace{3mm}
  \centerline{
   \epsfxsize8cm
 \epsffile{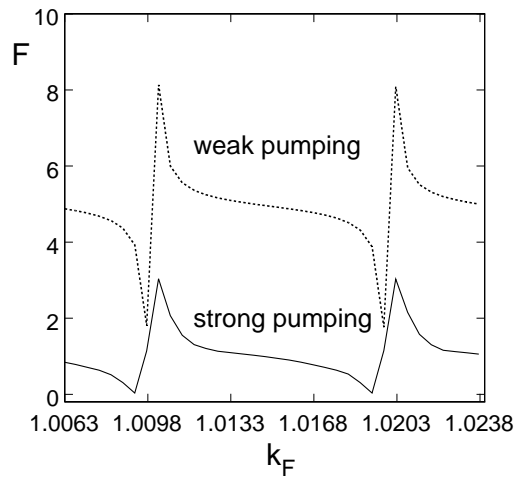}
             }
  \vspace{3mm}
  \nopagebreak
  \caption{
Current noise. The Fano factor for weak (dashed line)
and strong (solid line) pumping at zero temperature.
Parameters are the same as in Figs.\ref{fig1} and \ref{fig2}.
The transmission resonances through the double barrier structure are at 
$k_F=1.00968$ and at $k_F = 1.019677$.
   }
\label{fig3}
\end{figure}

To characterize the noise efficiency of the pump we use the Fano
factor $F$ [see, e.g., Ref.~\onlinecite{BB00}] which is the ratio
of the actual noise power ${\bf P}_{\alpha\alpha}$ and the Poisson
noise value ${\bf P}^{(P)}_{\alpha} = |2eI|$ corresponding to the
dc pumped current $I \equiv I_{1}$. We calculate the following
\begin{equation}
\label{Eq38}
F = \frac{{\bf P}_{11}}{|2eI_{1}|}
\end{equation}

In Fig.\ref{fig3} we present the Fano factor $F$  vs. $k_F$. Note
that qualitatively we find the same dependence for weak (dashed
line) and strong (solid line) pumping. However in the strong
pumping regime the Fano factor is smaller. Out of resonance the
Fano factor is close to unity. Close to resonance where the charge
pumped for a cycle $Q = I{\cal T} \approx e$ is approximately
quantized the Fano factor approaches zero that is fully consistent
with existing discussions \cite{AK00,AEGS01,MM01} in the
literature.

Note that the large value of the Fano factor, Eq. (\ref{Eq38}), in
the weak pumping regime has a simple explanation. 
In fact, the Fano factor, as defined in Eq. (\ref{Eq38}), does not reflect the
underlying physics of the pump effect. The working pump generates
non-equilibrium quasi-electron-hole pairs \cite{MB02} which
dissolve into quasi-electrons and quasi-holes after leaving the
scattering region. The entire pumped current $I$ consists of two
contribution: $I = I^{(e)} - I^{(h)}$. One of them $I^{(e)}$ is
due to the flow of quasi-electrons and the other $I^{(h)}$ is due
to the flow of quasi-holes. In the weak pumping regime we have
$I^{(e)}\approx I^{(h)}$ and $I \ll I^{(e)},I^{(h)}$. The noise is
a sum of statistically independent contributions from electrons
${\bf P}^{(e)}$ and holes ${\bf P}^{(h)}$ and a term which
describes correlations \cite{MB02} due to the common origin of
electron-hole pairs ${\bf P}^{(eh)}$. The noise power can be
expressed as a sum of these contributions: ${\bf P} = {\bf
P}^{(e)} + {\bf P}^{(h)} + {\bf P}^{(eh)}$. The Fano factor
calculated for each kind of particles, $F^{(x)} = {\bf
P}^{(x)}/(2eI^{(x)}), x = "e"," h"$ is close to unity. In
contrast, the Fano factor, Eq. (\ref{Eq38}), calculated with
respect to the total current $I$ can be much larger, reaching
super-Poissonian values.

\subsubsection{High temperature}
\label{SE_HT}

At high temperature the pump generates shot noise,
Eq. (\ref{Eq25}), and contributes to the thermal noise,
Eq. (\ref{Eq20C}). These two contributions depend on both the pump
frequency $\omega$ and the temperature $T$ in different ways.
Therefore we consider them separately.

For simplicity we suppose the temperature to be smaller then the
width $\delta E$ of the transmission resonance:
\begin{equation}
\label{Eq39}
\hbar\omega \ll k_BT \ll \delta E.
\end{equation}

In this case we can safely integrate over energy in
Eqs. (\ref{Eq20A}), (\ref{Eq25}) and can express the noise power in
terms of the scattering matrix elements given at the Fermi energy
$\mu$ . Performing the inverse Fourier transformation we find the
current auto correlations:
\begin{subequations}
\label{Eq40}
\begin{equation}
\label{Eq40A} {\bf P}_{\alpha\alpha}^{(sh)} = \frac{e^2\hbar}{6\pi
k_BT} \int\limits_{0}^{\cal T} \frac{dt}{\cal T}
\sum\limits_{\delta\neq\alpha} \left| \sum\limits_{\gamma}
\frac{\partial S_{\alpha\gamma}}{\partial t}  S_{\delta\gamma}^{*}
\right|^2
\end{equation}
\begin{equation}
\label{Eq40B} {\bf P}^{(th,p)}_{\alpha\alpha} =
i\frac{e^2k_BT}{\pi} \int\limits_{0}^{\cal T} \frac{dt}{\cal T}
\sum\limits_{\delta\neq\alpha} \left( \frac{\partial
S_{\alpha\delta}}{\partial t} \frac{\partial
S_{\alpha\delta}^{*}}{\partial E} - \frac{\partial
S_{\alpha\delta}}{\partial E} \frac{\partial
S_{\alpha\delta}^{*}}{\partial t} \right).
\end{equation}
\end{subequations}

\begin{figure}[t]
  \vspace{3mm}
  \centerline{
   \epsfxsize8cm
 \epsffile{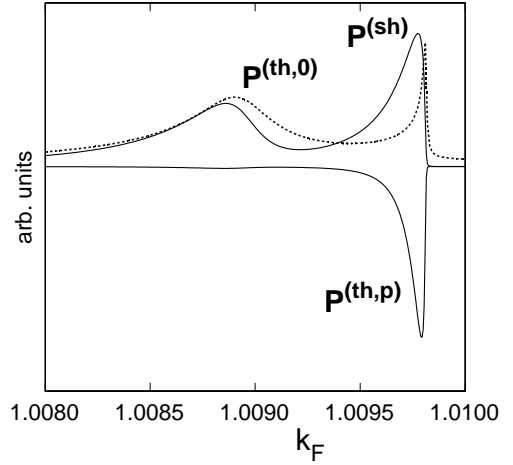}
             }
  \vspace{3mm}
  \nopagebreak
  \caption{
Current noise. High temperature strong pumping. The dependence of
the pump induced contributions to the noise power ${\bf
P}^{(sh)}\equiv{\bf P}^{(sh)}_{11}$, Eq. (\ref{Eq40A}), and ${\bf
P}^{(th,p)}\equiv{\bf P}^{(th,p)}_{11}$, Eq. (\ref{Eq40B}), on the
Fermi wave number $k_F$. For comparison the thermal noise ${\bf
P}^{(th,0)}\equiv{\bf P}^{(th,0)}_{11}$, Eq. (\ref{Eq20B}), (dashed
line) is presented as well. All the quantities are given in
arbitrary units, Parameters are the same as in Fig.\ref{fig2}.
   }
\label{fig4}
\end{figure}

In Fig. \ref{fig4} we depict ${\bf P}^{(th,p)}\equiv{\bf
P}^{(th,p)}_{11}$ and ${\bf P}^{(sh)}\equiv{\bf P}_{11}^{(sh)}$
(in arbitrary units) for strong pumping close to the transmission
resonance. The thermal noise ${\bf P}^{(th,0)} =
\frac{4e^2}{h}k_BT\big|S_{12}\big|^2 $ is depicted as well.

The behavior of the high temperature shot noise ${\bf P}^{(sh)}$
is similar to that of both the zero temperature shot noise,
Fig.\ref{fig2}, and the thermal noise ${\bf P}^{(th,0)}$,
Fig.\ref{fig4} (dashed line). In contrast, the pump induced
correction to the thermal noise
 ${\bf P}^{(th,p)}$ is essentially different.
In the case under consideration, first, it is essentially
negative, and, second, it contributes within a narrow energy
window.

\subsection{Heat flow noise power at zero temperature}

Since the heat flow noise differs from the current noise only at
low temperatures $k_BT \ll \hbar\omega$ we present the results of
calculations for the heat flow noise only for $T=0$.

Together with the heat flow noise we calculate the heat current
$I_{E,\alpha}$. In the adiabatic case the heat current reads:
\cite{MBstrong02}
\begin{equation}
\label{Eq41}
I_{E,\alpha} = \frac{\hbar\omega^2}{4\pi} 
\sum\limits_{\beta} \sum\limits_{q = -\infty}^{\infty}
q^2 \left| S_{\alpha\beta, q}(\mu)\right|^2.
\end{equation}

Like the current noise the heat flow noise is essentially
different in weak  ($V_{1j}\ll V_{0j}$) and strong ($V_{1j}\sim
V_{0j}$) pumping regimes.

\subsubsection{Weak pumping}

In this regime the heat flow noise peaks when the Fermi energy
$\mu$ matches a transmission resonance energy. This is very
similar to the current noise, Fig.\ref{fig1}. However  the
dependence on the phase difference $\varphi_1 - \varphi_2$ clearly
shows the difference between the heat flow noise and the current
noise. This is related to the fact that the pump is a source of
heat flows and thus neither the heat flows nor its fluctuations
are subject to the conservation laws, Eq. (\ref{Eq11}).

Note that the heat flows are sensitive to the spatial asymmetry of
the scatterer: if $V_{01}\neq V_{02}$ and/or $V_{11}\neq V_{12}$
then $I_{E,1}\neq I_{E,2}$. Similarly to the heat flow current,
the heat flow noise is sensitive to the spatial asymmetry of the
scatterer. In addition the heat flow auto correlations are
sensitive to a weak spatial asymmetry arising if the time reversal
invariance is broken by the working pump. \cite{MBstrong02} Let us
consider a stationary scatterer which is spatially symmetric:
$V_{01} = V_{02}$. Further suppose that the potentials $V_{01}$
and  $V_{02}$ oscillate with small amplitudes and with a phase lag
$\Delta\varphi \equiv \varphi_1 - \varphi_2$. If
$\Delta\varphi\neq 0$ then the time reversal invariance is broken.
This in turn breaks the spatial symmetry of a double barrier. Note
that neither the heat and charge currents nor the current noise do
feel this induced spatial asymmetry.

 \begin{figure}[t]
  \vspace{3mm}
  \centerline{
   \epsfxsize8cm
 \epsffile{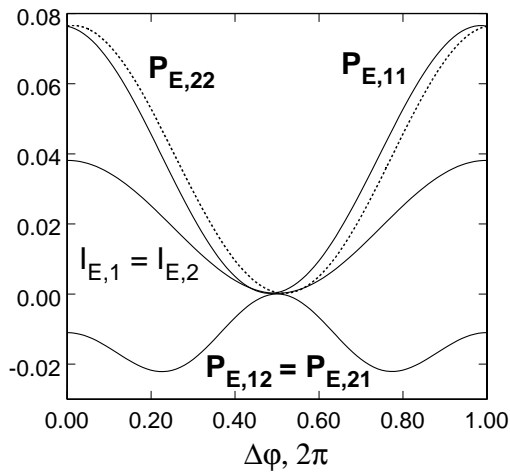}
             }
  \vspace{3mm}
  \nopagebreak
  \caption{
Heat flow noise at the transmission resonance. Zero temperature
weak pumping. The dependence of the dc heat flow
$I_{E,1}=I_{E,2}$, in units of $\hbar\omega^2/(4\pi)$, the heat
auto correlations ${\bf P}_{E,11}$, ${\bf P}_{E,22}$,  in units of
$10\hbar^2\omega^3/\pi$, and the heat cross correlations ${\bf
P}_{E,12} = {\bf P}_{E,21}$,  in units of $\hbar^2\omega^3/\pi$,
on the phase difference $\Delta\varphi = \varphi_1 - \varphi_2$,
in units of $2\pi$. $k_F = 1.00968$. The other parameters are the
same as in Fig.\ref{fig1}.
 }
\label{fig5}
\end{figure}

In Fig.\ref{fig5} we present the heat flow auto correlations ${\bf
P}_{E,jj}, j = 1,2$ and the heat flow cross correlations ${\bf
P}_{E,12}={\bf P}_{E,21}$ together with the dc heat flow
$I_{E,1}=I_{E,2}$ as a function of the phase difference $\varphi_1
- \varphi_2$ for the weak ($V_{1j} \ll V_{0j}$) pumping regime at
the transmission resonance. Note the different units for auto
correlations and cross correlations.

We see that the heat flow noise cross correlations are negative
like the current cross correlations. However unlike the current
noise the heat flow auto correlations are unrelated to the cross
correlations. In addition the heat flow auto correlations are
different at both leads. This is a consequence of the sensitivity to the
heat flow noise to the induced spatial asymmetry discussed above.

\subsubsection{Strong pumping}

\begin{figure}[t]
  \vspace{3mm}
  \centerline{
   \epsfxsize8cm
 \epsffile{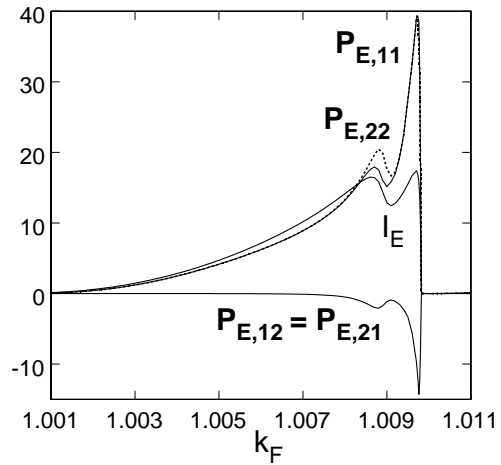}
             }
  \vspace{3mm}
  \nopagebreak
  \caption{
Heat flow noise. Zero temperature strong pumping. The dependence
of the dc heat flows $I_{E}\equiv I_{E,1}=I_{E,2}$, in units of
$\hbar\omega^2/(4\pi)$, and its noise ${\bf P}_{E,\alpha\beta}$,
in units of $100\hbar^2\omega^3/\pi$, on the Fermi wave number
$k_F$. The transmission resonance through the double barrier
structure appears at $k_F = 1.00968$. Parameters are the same as
in Fig.\ref{fig2}.
 }
\label{fig6}
\end{figure}

In Fig.\ref{fig6} we present the heat flow noise together with the
dc heat flow as a function of the Fermi wave number $k_F$ in the
vicinity of the transmission resonance at strong ($V_{1j} \sim
V_{0j}$) pumping. Note the increase of the ratio ${\bf
P}_{12}/{\bf P}_{\alpha\alpha}$ compared with the weak pumping
regime. In Fig.\ref{fig5} ${\bf P}_{12}$ and ${\bf
P}_{\alpha\alpha}$ are given on different scales whereas in
Fig.\ref{fig6} they are given on the same scale.

Close to the transmission resonance the heat noise and its
fluctuations show a non monotonic dependence on the Fermi energy
$\mu$ (the Fermi wave vector $k_F$). However unlike the current
noise, Fig.\ref{fig2}, the heat flow noise remains large even at
near quantized charge pumping.

\subsection{Breit-Wigner resonance}

We conclude  this section with a brief discussion of the sharp drop
of the pumped current and noise in the strong pumping regime
occurring just after the transmission resonance
[see, Fig.\ref{fig2}, Fig.\ref{fig4}, and Fig.\ref{fig6}].

At strong pumping the pumped charge is large if the pumping 
contour encloses the resonance line corresponding to  
$E = E_F$. \cite{LEWW01}
For our model, Eq. (\ref{Eq35}), and for well separated transmission 
resonances each time only one  resonance contributes to the pumped 
charge and noise.
If the Fermi energy $E_F$ changes then the corresponding resonance 
line [in the plane of pump parameters $(V_1,V_2)$] 
can leave the pumping contour
[the trace of the point with coordinates $(V_1(t), V_2(t))$ ]. 
This leads to a strong decrease of the pumped current.
To find the conditions when the resonance line is enclosed 
by the pumping cycle we proceed as follows. 

If the transmission resonances are well separated then only the one closest 
to the Fermi energy $E_F$ (say, $N$-th) 
contributes to the pumped current.
In such a case for our purposes we can approximate the scattering matrix 
$\hat S$ Eq. (\ref{Eq36})
by the Breit-Wigner scattering matrix $\hat S \approx \hat S_{BW}$: 
$$
 \hat S_{BW} = \frac{-1}{\Delta E+ i\frac{\Gamma}{2}}
\left(
\begin{array}{cc}
\Big( \Delta E + i\frac{\Gamma_2 - \Gamma_1}{2} \Big) e^{i\theta_1}
&  
i\sqrt{\Gamma_1 \Gamma_2} e^{i\theta} \\
\ \\  
i\sqrt{\Gamma_1 \Gamma_2} e^{i\theta}
&
\Big( \Delta E - i\frac{\Gamma_2 - \Gamma_1}{2} \Big) e^{i\theta_2}  
\end{array}
\right).
$$

\noindent The $N$-th  transmission resonance occurs at 
$E = E_{N}\equiv \frac{\hbar^2k_{N}^2}{2m}$ with 
$k_{N} \approx \frac{\pi N}{L}\Big( 1 - \frac{1}{2L}
\big( \frac{1}{p_1} + \frac{1}{p_2} \big) \Big)$. 
Here we introduced the following notations: 
$\Delta E = E - E_{N}$; $\Gamma =  \Gamma_1 + \Gamma_2$; 
$\Gamma_j = \frac{E_{N}k_{N}}{Lp_j^2}, j = 1,2$; 
$\theta = \pi N + \frac{1}{2}\big(\theta_1 + \theta_2 \big)$;
$\theta_j = \frac{k_{N}}{p_j}, j = 1,2$. 
Other parameters are defined after Eq. (\ref{Eq36}).

The pumping contour encloses the resonance line (at least, in part) 
if  the resonance condition $E_F = E_{N}$ 
can be achieved during the pumping cycle. 
Using the dependence $k_{N}(p_1,p_2)$ 
we rewrite the resonance condition as follows ($p_j = mV_j/\hbar^2$):
$$
\frac{1}{V_1(t)} + \frac{1}{V_2(t)} = 
\frac{mL}{\hbar^2 E_{N0}}\Big( E_{N0} - E_F \Big),
$$
where 
$E_{N0} =  \frac{\hbar^2}{2m}\big( \frac{\pi N}{L} \big)^2$.
Since all the quantities are positive one can conclude that 
the above equation has a solution only if $E_F < E_{N0}$. 
In contrast, at $E_F > E_{N0}$ the resonance line lies fully away from 
the pumping cycle and the pumped charge is vanishingly small.
Therefore with increasing $E_F$ the pumped current (and noise) decreases 
as soon as the Fermi energy crosses some resonance state 
$E_{N0}$  in a nearly isolated ($V_j \to \infty$) well.

\section{Conclusion}
\label{DC}

We have generalized the scattering matrix approach to adiabatic
pumping in a system of noninteracting spinless fermions 
of Ref.~\onlinecite{MB02} to the case of a strong amplitude quantum pump.
Consequently, our
approach takes into account the energy absorption and
emission (by modulation quanta $n\hbar\omega,~n = 1,2,\dots $) of
electrons traversing the oscillating scatterer. This allows a
description of the correlation properties (noise) of a pump at
high ($k_BT\gg\hbar\omega$) as well as at low
($k_BT\ll\hbar\omega$) temperatures. 
We found that the noise produced by the pump is described by photon-assisted
quantum mechanical exchange amplitudes.

At low temperature the pump is the main
source of a noise. The low temperature noise is linear in pump
frequency $\omega$ and it can be viewed as an analog of the shot
noise in dc-biased conductors. \cite{BB00} At high temperature the
pump produces characteristic high temperature corrections to the
equilibrium Nyquist-Johnson noise. These corrections are quadratic
in the pump frequency  at lower temperatures but become linear in
$\omega$ if the temperature increases, Eq. (\ref{Eq29}). The last
contribution is due to existence of instantaneous  currents,
Eqs. (\ref{Eq21}) and (\ref{Eq22}), generated by the pump.

The pump is a source of heat flows. \cite{MB02} 
From the scatterer to the reservoirs the energy is
carried by electrons traversing the pump. 
The quantum statistical nature of electron flow leading 
to current noise results in heat flow fluctuations as well. 
At low temperature ($k_BT\ll\hbar\omega$) the pump produces a heat flow 
shot noise which exceeds the equilibrium Nyquist-Johnson heat flow noise. 

The theory presented allows a unified
description of correlation properties of multi-terminal quantum
pumps serving as a source of both current and energy flows
at low as well as at relatively high temperatures. We argue that even the
adiabatic quantum pump is at zero
temperature a far from equilibrium system
but it approaches a quasi-equilibrium state as the
temperature increases beyond the scale $\hbar\omega$ determined by
the pump frequency $\omega$. We found that at low temperature the
correlation properties of energy flows differ essentially from
the correlation properties of currents generated by the pump.
In contrast at high temperature both correlations are related to each
other in accordance with the fluctuation dissipation theorem.

\begin{acknowledgments}

We thank Peter Samuelsson for an important comment.
M.M. appreciates the warm hospitality of the Department of
Theoretical Physics of the University of Geneva, where part of
this work was done.

This work was supported by the Swiss National Science
Foundation and the EU RTN under Contract N0. HPRN-CT-2000-00144.

\end{acknowledgments}

\appendix*
\section{}

\subsection{Linear in frequency thermal noise}
\label{A0}

In this section we present in detail the calculation of the thermal noise, 
Eq. (\ref{Eq10B}), within the adiabatic approximation.
First, using the adiabatic approximation, Eq. (\ref{Eq16}), 
we evaluate the transmission probabilities from the
Floquet scattering matrix element with an accuracy of $\omega$:
$$
\begin{array}{c}
\big| S_{F,\alpha\gamma}(E_n,E) \big|^2 =
\left( 1 + \frac{n\hbar\omega}{2}\frac{\partial}{\partial E} \right)
\big| S_{\alpha\gamma,n}(E) \big|^2 \\
\ \\
+ 2\hbar\omega Re [S_{\alpha\gamma,n}^{*}(E)
A_{\alpha\gamma,n}(E)] + O(\omega^2).
\end{array}
$$
Then performing the inverse Fourier transformation we sum over $n$:
$$
\begin{array}{c}
\sum\limits_{n=-\infty}^{\infty} 
\big| S_{F,\alpha\gamma}(E_n,E) \big|^2 =
\int\limits_{0}^{\cal T}\frac{dt}{\cal T} \bigg\{
\big| S_{\alpha\gamma}(E,t) \big|^2 \\
\ \\
+ \frac{i\hbar}{2}\Big(
\frac{\partial S_{\alpha\gamma}(E,t)}{\partial t}
\frac{\partial S_{\alpha\gamma}^{*}(E,t)}{\partial E}
+ \frac{\partial^2 S_{\alpha\gamma}}{\partial E\partial t} 
S_{\alpha\gamma}^{*}(E,t) \Big) \\
\ \\
+ 2\hbar\omega Re [S_{\alpha\gamma}^{*}(E,t) 
A_{\alpha\gamma}(E,t)] \bigg\} + O(\omega^2).
\end{array}
$$
To remove the second order derivative we integrate by parts over $t$.
Further using an equality which can be obtained from Eq. (\ref{Eq18})
(see also Ref.\onlinecite{MBac03})
$$
4\hbar\omega \sum\limits_{\gamma} Re[S^{*}_{0,\alpha\gamma}
A_{\alpha\gamma}] = 
{\cal P}\{\hat S_{0};\hat S_{0}^{\dagger} \}_{\alpha\alpha},
$$
we sum over $\gamma$:
$$
\begin{array}{c}
\sum\limits_{\gamma}\sum\limits_{n=-\infty}^{\infty} 
\big| S_{F,\alpha\gamma}(E_n,E) \big|^2 =  \\
\ \\
1 + i\hbar \int\limits_{0}^{\cal T}\frac{dt}{\cal T}
\Big(
 \frac{\partial\hat S(E,t)}{\partial t}
\frac{\partial\hat S^{\dagger}(E,t)}{\partial E}
- \frac{\partial\hat S(E,t)}{\partial E}
\frac{\partial\hat S^{\dagger}(E,t)}{\partial t}
\Big)_{\alpha\alpha} + O(\omega^2).
\end{array}
$$
The remaining part is as follows
$$
\begin{array}{c}
- \sum\limits_{n=-\infty}^{\infty} \bigg (
\big| S_{F,\alpha\beta}(E_n,E) \big|^2
+ \big| S_{F,\beta\alpha}(E_n,E) \big|^2
\bigg)  = \\
\ \\
 - \int\limits_{0}^{\cal T}\frac{dt}{\cal T} \bigg\{
\big| S_{\alpha\beta}(E,t) \big|^2 + \big| S_{\beta\alpha}(E,t) \big|^2 
+ \frac{h}{e}\frac{dI_{\alpha\beta}^{(s)}(E,t)}{dE} \bigg\} 

+ O(\omega^2).
\end{array}
$$
Here we have introduced a symmetrized spectral current density 
$dI_{\alpha\beta}^{(s)}/dE = dI_{\alpha\beta}/dE + dI_{\beta\alpha}/dE$
generated by the pump. \cite{MBac03}
The spectral current density is: 
$$
\frac{dI_{\alpha\beta}}{dE} = \frac{e}{h}\Big(
2\hbar\omega Re[S^{*}_{0,\alpha\beta}A_{\alpha\beta}]
+\frac{1}{2}{\cal P}\{S_{0,\alpha\beta}; S^{*}_{0,\alpha\beta}\} \Big). 
$$

Finally collecting together all the terms we obtain 
Eqs. (\ref{Eq20}).

\subsection{Two-particle scattering matrix }
\label{A1}

The scattering matrix $\hat S$ collects all the quantum mechanical 
amplitudes corresponding to the scattering of a single particle 
by the multi-terminal sample.
This quantity is appropriate to describe the observable, like a current,
which can be represented as a sum over single-particle states.

The noise, the quantity of interest in the present paper, is different.
It is a current-current correlation function and,
thus, the noise involves essentially two-particle scattering
(even for noninteracting particles). \cite{BB00}
To characterize two-particle scattering it is convenient to introduce
{\it a two-particle scattering matrix} $\hat \Sigma$.

In a general non-stationary case the scattering matrix $\hat S$ 
depends on two times
(an arrival time, and a departure time), and thus the matrix
$\hat\Sigma$ depends on four times (two times for each particle).
However within the adiabatic picture used in the present paper 
the scattering matrix $\hat S(t)$ depends on only one time. 
It is the scattering matrix {\it frozen} 
at the time the particle traverses the scatterer.
Therefore within the frozen scattering matrix approximation 
the matrix $\hat \Sigma$ depends on two times only.
We define this matrix as follows:
\begin{equation}
\label{EqA_1}
\hat\Sigma(t_1,t_2) = \hat S(t_1) \hat S^{\dagger}(t_2).
\end{equation}
This definition is appropriate for describing of 
an electron hole scattering process 
relevant for the zero temperature shot noise, Sec.\ref{ANP_ZT}.
The time arguments determine the scattering
matrix encountered by the electron at time 
$t_1$ and by the hole at time $t_2$.

The scattering matrix Eq. (\ref{EqA_1}) assumes implicitly that 
electron and hole enter the scatterer through the same lead.
Note if an electron and hole are in the same lead then they have no effect on
the average current nor on the zero frequency noise of interest here. 
Therefore we can say that 
the scattering matrix $\hat\Sigma$ describes {\it effectively} 
a two particle scattering process with only an outgoing electron and a hole.
These outgoing particles can be viewed as composing 
an electron-hole pair generated by the pump.

In this Appendix we collect some evident but useful properties 
of a two-particle frozen scattering matrix $\hat\Sigma(t_1,t_2)$.

The matrix $\hat\Sigma(t_1,t_2)$ is unitary
\begin{equation}
\label{EqA_2}
\hat\Sigma(t_1,t_2)\hat\Sigma^{\dagger} (t_1,t_2) = \hat I,
\end{equation}
\noindent
and it is a unit matrix at coincident times
\begin{equation}
\label{EqA_3}
\hat\Sigma(t_1,t_1) = \hat I.
\end{equation}
\noindent
Here $\hat I$ is a unit matrix with matrix elements 
$I_{\alpha\beta} = \delta_{\alpha\beta}$.

At times which are close to each other:
$t_1= t + \Delta t$,  $t_2 = t$, $\Delta t\to 0$, we can write
\begin{equation}
\label{EqA_4}
\hat\Sigma(t_1, t) = \hat I + \big(t_1 - t \big) \partial_{t_1}\hat\Sigma(t)
+ \frac{(t_1 - t)^2}{2} \partial^2_{t_1}\hat\Sigma(t) + \dots.
\end{equation}
\noindent
Here we have introduced the abbreviations
$\partial^n_{t_1}\hat\Sigma(t) =
\frac{\partial^n\hat\Sigma(t_1,t)}{\partial t_1^n}\Big|_{t_1=t}
\equiv \frac{\partial^n \hat S(t)}{\partial t^n} \hat S^{\dagger}(t)$.
The diagonal element
$\partial_{t_1}\hat\Sigma_{\alpha\alpha}(t) = 
\frac{-2\pi i}{e}I_{\alpha}(t)$
is proportional to a zero temperature current 
$I_{\alpha}$ pushed by the scatterer into the lead $\alpha$.

For the scatterer with periodically evolving properties the matrix
$\hat \Sigma$ is periodic in both arguments:
\begin{equation}
\label{EqA_5}
\hat \Sigma(t_1,t_2) = \hat \Sigma(t_1+{\cal T},t_2) = 
\hat \Sigma(t_1,t_2+{\cal T}).
\end{equation}

The squared matrix element $|\Sigma_{\alpha\beta}(t_1,t_2)|^2$
is the probability for an electron and a hole to leave the scattering region 
through the lead $\alpha$  at time moment $t_1$
and the lead $\beta$ at time moment $t_2$, respectively.
Averaging it over the middle time $t = \frac{t_1+t_2}{2}$
we find the  matrix element of the matrix
$\hat C(\tau)$ (where $\tau = t_1 - t_2$)
which defines the correlations of currents produced by the pump:
\begin{equation}
\label{EqA_6}
C_{\alpha\beta}(\tau) = \int\limits_{0}^{\cal T}\frac{dt}{\cal T}
\left|\Sigma_{\alpha\beta}\left(t 
+ \frac{\tau}{2},t - \frac{\tau}{2} \right) \right|^2.
\end{equation}

By definition the matrix $\hat C(\tau)$ is real and positive.
In addition, since the matrix $\hat \Sigma$ is unitary
we have
\begin{equation}
\label{EqA_7}
\sum\limits_{\alpha} C_{\alpha\beta}(\tau) = 
\sum\limits_{\beta} C_{\alpha\beta}(\tau) = 1.
\end{equation}

The matrix $\hat C_q$ of the corresponding Fourier coefficients
(entering the equation (\ref{Eq24}) for the noise)
\begin{equation}
\label{EqA_8}
\hat C_q = \int\limits_{0}^{\cal T} 
\frac{d\tau}{\cal T} e^{iq\omega \tau} \hat C(\tau),
\end{equation}
\noindent
is a self adjoint matrix
\begin{equation}
\label{EqA_9}
\hat C_{q} = \hat C_{q}^{\dagger}.
\end{equation}

This equation can be easily proven if one expresses 
$C_{\alpha\beta, q}$ in terms of the
(Fourier coefficients for the product of  two)
elements of the single particle scattering matrix  $\hat S$
[Eqs. (\ref{EqA_1}), (\ref{EqA_6}) and (\ref{EqA_8})]:
\begin{equation}
\label{EqA_10}
C_{\alpha\beta,q} = \sum\limits_{\gamma}\sum\limits_{\delta}
\left( S_{\alpha\gamma}^{*}S_{\alpha\delta} \right)_{q}
\left( S_{\beta\gamma}^{*}S_{\beta\delta} \right)_{q}^{*}.
\end{equation}
\noindent
Here we used: $A^{*}_{q} = (A^{*})_{-q}$.

Equation (\ref{EqA_10}) leads to the relation:
$C_{\alpha\beta,q} = C_{\beta\alpha,q}^{*}$, 
which is nothing but Eq. (\ref{EqA_9}).
In addition, since $C_{\alpha\beta}$ is real and therefore
$C_{\alpha\beta,q} = C_{\alpha\beta,-q}^{*}$,
we have  $C_{\alpha\beta,-q}^{*} = C_{\beta\alpha,q}^{*}$.
This means that the matrix $\hat C(\tau)$ is subject to the following
symmetry condition
\begin{equation}
\label{EqA_11}
C_{\alpha\beta}(\tau) =  C_{\beta\alpha}(-\tau).
\end{equation}
We would like to emphasize that this
condition has nothing to do with a true micro reversibility condition.
In a general non-stationary case \cite{MBstrong02} the micro 
reversibility implies not only a reversal of the time argument and
an interchange of lead indices
(or more generally, interchanging of incoming and outgoing channels)
but also the change of the relative phases of all relevant time-dependent 
processes. For instance, if two parameters
$X_i = X_{0i}\cos(\omega t + \varphi_i),~ i = 1,2$
of a scatterer change periodically in time with phase lag
$\Delta\varphi = \varphi_1 - \varphi_2 \neq 0$
then the micro reversibility means 
(in an adiabatic case) 
an invariance under the following interchange:
$ (\alpha,\beta, t, \Delta\varphi) \to (\beta,\alpha, -t, -\Delta\varphi) $.
In contrast,  the matrix $\hat C(\tau)$ is invariant already under 
the substitution: $ (\alpha,\beta, \tau) \to (\beta,\alpha, -\tau) $
(note that $\tau=t_1-t_2$ is a time difference which obviously changes a sign
when time is reversed).

If time reversal invariance (TRI) is present:
$\hat C^{(TRI)}(\tau) = \hat C^{(TRI)}(-\tau)$,
then it follows from Eq. (\ref{EqA_11}) that the matrix 
$\hat C^{(TRI)}$ is symmetric in lead indices
\begin{equation}
\label{EqA_12}
C^{(TRI)}_{\alpha\beta} =  C^{(TRI)}_{\beta\alpha}.
\end{equation}

Note the symmetrized matrix
$\hat C^{(sym)}(\tau) = \frac{1}{2}\left( \hat C(\tau) + \hat C(-\tau)\right)$
defining the zero frequency shot noise power, Eq. (\ref{Eq28}),
is obviously symmetric in lead indices as well
\begin{equation}
\label{EqA_13}
C^{(sym)}_{\alpha\beta} =  C^{(sym)}_{\beta\alpha}.
\end{equation}

Finally note the useful equality
\begin{equation}
\label{EqA_14}
\sum\limits_{q=-\infty}^{\infty} \hat C_{q} = \hat I.
\end{equation}
\noindent
The proof is as follows.
Using Eq. (\ref{EqA_10}) we get
$$
\begin{array}{c}
\sum\limits_{q=-\infty}^{\infty}C_{\alpha\beta,q} =
\sum\limits_{\gamma}\sum\limits_{\delta}
\sum\limits_{q=-\infty}^{\infty}
\left( S_{\alpha\gamma}^{*}S_{\alpha\delta} \right)_{q}
\left( S_{\beta\gamma}^{*}S_{\beta\delta} \right)_{q}^{*} \\
\ \\
= \int\limits_{0}^{\cal T} \frac{d\tau}{\cal T}
\sum\limits_{\gamma} S_{\alpha\gamma}^{*}(\tau)S_{\beta\gamma}(\tau)
\sum\limits_{\delta} S_{\alpha\delta}(\tau)S_{\beta\delta}^{*}(\tau) 
= \delta_{\alpha\beta}.
\end{array}
$$
Here we used the unitarity condition for the single particle 
scattering matrix.

\subsection{The matrix $\hat D$}
\label{A2}

We define the matrix $\hat D(\tau)$,
whose Fourier coefficients enter the expression for the zero temperature
heat  noise power, Eq. (\ref{Eq32}),
in analogy with the matrix $\hat C(\tau)$, Eq. (\ref{EqA_6}):
\begin{equation} 
\label{EqA_15}
D_{\alpha\beta}(t_1-t_2) = 
\frac{1}{4}\int\limits_{0}^{\cal T}\frac{d(t_1+t_2)}{\cal T}
\Big( \Sigma_{\alpha\beta}(t_1,t_2) 
\Xi^{*}_{\alpha\beta}(t_1,t_2) + c.c. \Big).
\end{equation}
\noindent Here $c.c.$ means complex conjugate terms.
Here we have introduced a new matrix:
\begin{equation}
\label{EqA_16}
\hat \Xi(t_1,t_2) = \hbar^2 \frac{\partial \hat S(t_1)}{\partial t_1}
\frac{\partial \hat S^{\dagger}(t_2)}{\partial t_2}.
\end{equation}
This matrix can be represented as a product of
two energy shift matrixes and a two-particle scattering matrix:
\begin{equation}
\label{EqA_17}
\hat \Xi(t_1,t_2) = \hat{\cal E}(t_1) 
\hat\Sigma(t_1,t_2)\hat{\cal E}^{\dagger}(t_2).
\end{equation}

The energy shift matrix $\hat {\cal E}(t)$
is defined as follows: \cite{AEGS01,AEGS03}
$$
\hat{\cal E}(t) = 
i\hbar\frac{\partial \hat S(t)}{\partial t}\hat S^{\dagger}(t).
$$

The matrix $\hat \Xi(t_1,t_2)$ emphasizes an importance of energies of
particles for the heat flow fluctuations.

\end{document}